\documentclass[twocolumn,tighten]{aastex6}

\usepackage{placeins}
\usepackage{natbib}
\usepackage{amsmath}
\usepackage{color}

\begin{document}

\newcommand\be{\begin{equation}}
\newcommand\en{\end{equation}}
\newcommand\pdv{$P{\rm d}V~$}
\newcommand\brunt{Brunt-V\"ais\"al\"a }

\shorttitle{A Planet-Driven Spiral Wave Instability} 
\shortauthors{Bae et al.}

\title{THE SPIRAL WAVE INSTABILITY INDUCED BY A GIANT PLANET: I. PARTICLE STIRRING IN THE INNER REGIONS OF PROTOPLANETARY DISKS}

\author{Jaehan Bae\altaffilmark{1},
Richard P. Nelson\altaffilmark{2},
Lee Hartmann\altaffilmark{1}}

\altaffiltext{1}{Department of Astronomy, University of Michigan, 1085 S. University Ave.,
Ann Arbor, MI 48109, USA} 
\altaffiltext{2}{Astronomy Unit, Queen Mary University of London, Mile End Road, London E1 4NS, UK}

\email{jaehbae@umich.edu, r.p.nelson@qmul.ac.uk, lhartm@umich.edu}

\begin{abstract}

We have recently shown that spiral density waves propagating in accretion disks can undergo a parametric instability by resonantly coupling with and transferring energy into pairs of inertial waves (or inertial-gravity waves when buoyancy is important).  In this paper, we perform inviscid three-dimensional global hydrodynamic simulations to examine the growth and consequence of this instability operating on the spiral waves driven by a Jupiter-mass planet in a protoplanetary disk. We find that the spiral waves are destabilized via the spiral wave instability (SWI), generating hydrodynamic turbulence and sustained radially-alternating vertical flows that appear to be associated with long wavelength inertial modes. In the interval $0.3~R_p\leq~R\leq~0.7R_p$, where $R_p$ denotes the semi-major axis of the planetary orbit (assumed to be 5~au), the estimated vertical diffusion rate associated with the turbulence is characterized by $\alpha_{\rm diff}\sim~(0.2-1.2)~\times~10^{-2}$. For the disk model considered here, the diffusion rate is such that particles with sizes up to several centimeters are vertically mixed within the first pressure scale height. This suggests that the instability of spiral waves launched by a giant planet can significantly disperse solid particles and trace chemical species from the midplane.  In planet formation models where the continuous local production of chondrules/pebbles occurs over Myr time scales to provide a feedstock for pebble accretion onto these bodies, this stirring of solid particles may add a time constraint: planetary embryos and large asteroids have to form before a gas giant forms in the outer disk, otherwise the SWI will significantly decrease the chondrule/pebble accretion efficiency.

\end{abstract}


\section{INTRODUCTION}

When they form, planets leave traces of their presence in the disks they reside in.
The most suggestive signatures include spiral arms, of which we might already have observational snapshots \citep[e.g.,][]{muto12,garufi13,grady13,currie14,benisty15,garufi16,stolker16}, though the physical origin of the observed spiral arms as well as the presence of planet(s) in the systems are yet to be confirmed.

In our recent work, we have shown that propagating spiral density waves, such as the ones that could be excited by a planet embedded in a gaseous protoplanetary disk, or by gravitational instability (GI), can become unstable to a spiral wave instability (SWI; \citealt{bae16}).
The instability arises because the periodic forcing due to the spiral waves resonantly couples with, and transfers energy into, pairs of inertial waves (or inertial-gravity waves) at the expense of the spiral wave itself, such that the spiral waves partially dissipate.
When the instability tends towards nonlinear saturation, the flow breaks down into hydrodynamic turbulence, which in turn can act as an efficient source of vertical mixing of trace chemical species and solid particles \citep{bae16}.

The level of turbulence in protoplanetary disks plays an important role in determining the ability of solid particles to grow from ISM sizes to eventually become planets.
Solid particles grow to millimeters to centimeters in size through direct sticking collisions, after which point the so-called ``bouncing barrier'' limits further growth by hit-and-stick coagulation (see review by \citealt{testi14} and references therein). 
If decimeter-sized particles are able to form in abundance then streaming instabilities can concentrate these ``pebbles" and ``boulders" via aerodynamic drag, and the resulting particle clumps can collapse gravitationally to form planetesimals of between 25 to 200~km in size \citep{youdin05,johansen07,johansen15}.
Once large planetesimals are present, gas drag acting on the remaining pebbles can produce rapid accretion -- pebble accretion \citep{johansen10,ormel10,lambrechts12,morbidelli12}.
Pebble accretion involving millimeter-sized chondrules can help to build up planetary embryos to the size of Mars, and can also help to grow 100~km sized asteroids that form as a result of streaming instabilities into significantly larger bodies that resemble the largest asteroids in the asteroid belt \citep[e.g.,][]{johansen15}.
While the former process -- the streaming instability -- has been shown to operate in the presence of moderately strong turbulence with $\alpha \sim 10^{-3}$ \citep{johansen07}, where $\alpha$ denotes the canonical \citet{shakura73} stress parameter, for the latter process -- pebble accretion -- to work efficiently, it is necessary to concentrate pebbles into a thin layer at the disk midplane such that the scale height of the pebbles is smaller than the Hill radius of pebble-accreting planetesimals \citep{lambrechts12}.
Under typical disk conditions, this requires the scale height of pebbles to be less than $\sim1~\%$ of that of gas, which can only be satisfied with very low turbulence to avoid stirring up the pebbles.  
A turbulence level of $\alpha \gtrsim 10^{-4}$ may result in a too long timescale for the formation of planetary embryos or large asteroids \citep[e.g.,][]{johansen15}.

In the present paper, we perform three-dimensional global hydrodynamic simulations to demonstrate that the SWI operates in the presence of spiral waves excited by a Jovian mass giant planet, and to estimate the level of SWI-induced turbulence and the consequent vertical mixing of solid particles in the terrestrial body-forming and asteroid belt regions. 
Our simulations show that the SWI develops for the spiral waves on timescales on the order of the planetary orbital time.
When the instability is fully saturated, the vertical diffusion rate estimated from gas motions in the interval $0.3 R_p \leq R \leq 0.7 R_p$, where $R_p$ is the semi-major axis of the planetary orbit (assumed to be 5~au in this work), is such that particles with sizes up to a few centimeters are vertically mixed within the first pressure scale height of the gas disk.
This result suggests that the instability acting on the spiral waves from a gas giant can have significant influence on the dust dynamics and chemical mixing in protoplanetary disks.
In particular, if accretion of chondrules/pebbles dominates the growth of terrestrial embryos in the terrestrial planet region, or large asteroids in the asteroid belt, then we suggest that the SWI can have a strong influence by limiting chondrule/pebble accretion efficiency when a gas giant planet such as Jupiter forms in the outer disk.

This paper is organized as follows. 
In Section \ref{sec:expectation}, we discuss the expected wavelengths of the unstable inertial waves and the regions of a disk in which the SWI operates, based on the WKBJ dispersion relations.
We describe our computational setup in Section \ref{sec:method}, with a special emphasis on the choice of numerical resolution that suits for capturing unstable inertial modes.
In Section \ref{sec:results}, we begin by describing numerical results obtained with perturbations having {\it monochromatic} azimuthal modes with $m=2$ and $m=3$, with which the spiral arms are evenly spaced in azimuth. 
We then describe results obtained with a full planetary potential.
We present the analysis on the vertical mixing of particles arising from the SWI and discuss implications to the formation and growth of terrestrial bodies and large asteroids in Section \ref{sec:discussion}, and draw conclusions in Section \ref{sec:conclusion}.

\section{THEORETICAL EXPECTATIONS}
\label{sec:expectation}

\subsection{Monochromatic spiral waves}
\label{sec:monochromatic}
Before we consider spiral density waves launched by planets, which consist of a superposition of various azimuthal components \citep{goldreich79}, we briefly discuss theoretical expectation for the SWI driven by monochromatic $m=2$ and $m=3$ spiral waves to gain insight into the instability. 
These modes are chosen because they appear prominently in our simulations with Jupiter-mass planets, as discussed below. We focus mainly on spiral waves that propagate in the disk regions that lie interior to the planet in this paper, and leave discussion of outward propagating spiral waves for future work.
Throughout the paper, we use the term {\it monochromatic modes/waves} to denote the spiral waves that are evenly spaced in azimuth.
For a more detailed discussion of the theoretical background, we refer readers to Section 2 of \citet{bae16} and references therein.

In a differentially rotating disk, the WKBJ dispersion relation for local disturbances can be written as \citep{goodman93}
\begin{equation}
\frac{\omega^2 /c_s^2}{\omega^2 - N^2} - \frac{k_ Z^2}{\omega^2 - N^2} - \frac{k_R^2}{\omega^2 - \kappa^2} = 0.
\label{eqn:dispersion1}
\end{equation}
In the dispersion relation, $\omega$ is the mode frequency, $\kappa$ is the epicyclic frequency,
$N$ is the vertical Brunt-V\"ais\"al\"a frequency, $k_R$ and $k_Z$ are the radial and vertical wave numbers associated with the wave vector ${\bf k} = k_R {\bf {\hat e}}_R + k_Z {\bf {\hat e}}_Z$, and $c_s$ is the sound speed. 

The WKBJ dispersion relation for a spiral wave with azimuthal mode number $m$ is given by
\begin{equation}
m^2 (\Omega - \Omega_{\rm p})^2 =\kappa^2 + c_s^2 k_{R,{\rm s}}^2,
\label{eqn:spiral-dispersion}
\end{equation}
when self-gravity is neglected and $k_{Z,{\rm s}}=0$ is assumed.
Interior to the inner Lindblad resonance (ILR), this gives a radial wave number $k_{R,{\rm s}}$ of 
\begin{equation}
k_{R, {\rm s}}^2 = \frac{\omega_s^2 - \kappa^2}{c_s^2},
\label{eqn:k-spiral}
\end{equation}
where $\omega_s = m(\Omega-\Omega_p)$ is the Doppler-shifted frequency of the incoming waves, and $\Omega_{\rm p}$ is the pattern speed of the spiral wave measured in the inertial frame. 

Inertial modes excited by a spiral wave must have radial length scales that are similar to or smaller than the wavelength of the incoming wave.
Thus, assuming that the excited inertial modes have very similar spatial structure and frequencies, which is appropriate to the high wave number limit, we can relate the wave number of the excited inertial waves $k_{R, {\rm i}}$ to $k_{R, {\rm s}}$ as
\be
\label{eqn:kRi}
k_{R, {\rm i} }= n k_{R, {\rm s}},
\en
where $n$ is an integer. 
As noted in \citet{bae16}, this integer relationship holds for waves in periodic shearing boxes, and is introduced for the global models that we consider here for the purpose of choosing a simple relation between the wavelengths of the spiral and inertial waves.
Given the radial wave number and the frequency of the spiral wave (in the local fluid frame), we can estimate the vertical wave number of the excited inertial waves $k_{Z,{\rm i}}$ using the dispersion relation in Equation~(\ref{eqn:dispersion1}) and 
the resonance condition $\omega_{\rm i}=\omega_{\rm s}/2$ in the high wave number limit \citep[$n \gg 1$;][]{fromang07}.
This results in 
\begin{equation}
k_{Z,{\rm i}}^2 = \frac{(\omega_{\rm s}/2)^2}{c_s^2} - n^2 \frac{\omega_{\rm s}^2- \kappa^2}{c_s^2} \frac{(\omega_{\rm s}/2)^2 - N^2}{(\omega_{\rm s}/2)^2 - \kappa^2},
\label{eqn:kZi2}
\end{equation}
when the \brunt frequency, $N$, is important or
\begin{equation}
k_{Z,{\rm i}}^2 = \frac{(\omega_{\rm s}/2)^2}{c_s^2} \left(1 - n^2 \frac{\omega_{\rm s}^2-\kappa^2}{(\omega_{\rm s}/2)^2 - \kappa^2} \right)
\label{eqn:kZi3}
\end{equation}
when $N$ is negligible.
In order for $k_{Z,{\rm i}}$ to be real for arbitrary values of $n$ from Equation (\ref{eqn:kZi3}), such that inertial waves can be excited, one requires 
\be
\label{eqn:res_condition}
(\omega_{\rm s}/2)^2 < \kappa^2 < \omega_{\rm s}^2
\en
to be satisfied.

In rotating disks, inertial modes have frequencies in the interval $0 \leq \omega_i \leq \Omega$.
This, together with Equation (\ref{eqn:res_condition}), implies that for an arbitrary perturbation with azimuthal mode number $m \geq 1$ the SWI operates in the radial region of 
\be
\left( {m-2 \over m} \right)^{2/3} R_p \leq R \leq \left( {m-1 \over m} \right)^{2/3} R_p
\en
in the inner disk and 
\be
\left( {m+1 \over m} \right)^{2/3} R_p \leq R \leq \left( {m+2 \over m} \right)^{2/3} R_p
\en
in the outer disk.
Focusing on the inner disk, there will always be inertial mode pairs available for a resonant interaction with $m=2$ spiral waves having a Doppler-shifted frequency $\omega_s = 2(\Omega-\Omega_p) < \Omega$ inside the ILR.
On the other hand, for $m \geq 3$ waves, the SWI cannot operate in disk regions where $R \leq ((m-2)/m)^{2/3} R_p$, because the spiral wave frequency at these radii is too large, so no inertial modes can participate in a resonant interaction.

%figure 1
\begin{figure}
\centering
\epsscale{1.0}
\plotone{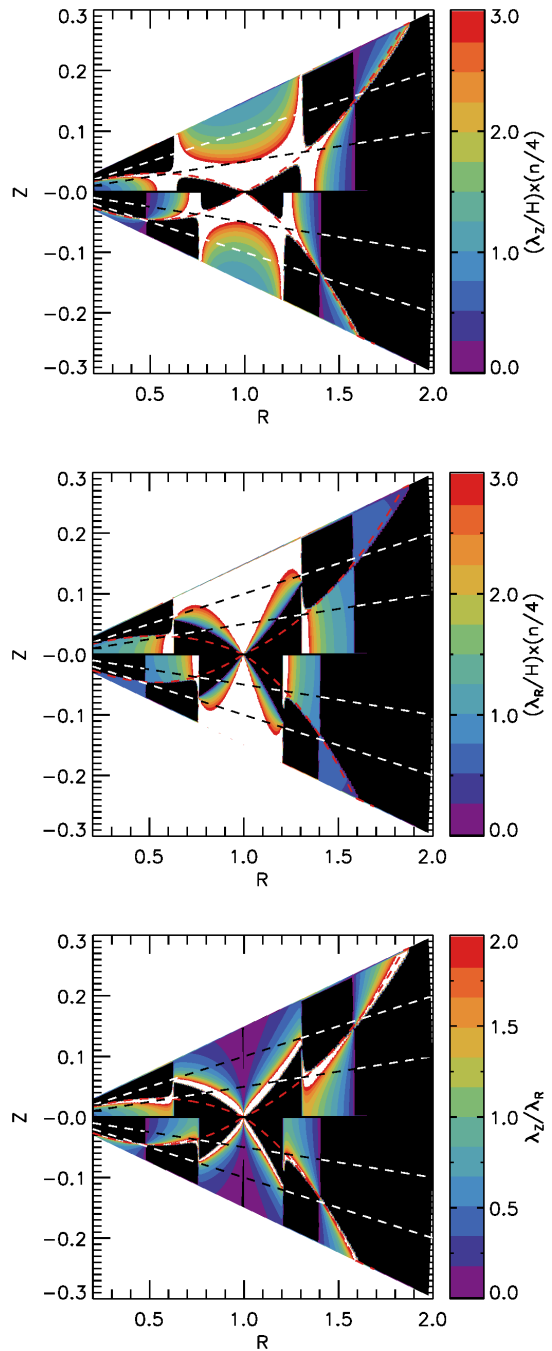}
\caption{(a) Contour plots of the vertical wavelength of the unstable inertial modes $\lambda_Z$, in units of local scale height, calculated with Equation (\ref{eqn:kZi2}) and the disk model that will be introduced in Section \ref{sec:model}. (b) Same as (a) but for the radial wavelength $\lambda_R$. (c) The ratio of vertical to radial wavelength of the inertial modes. In all panels, the black and white dashed lines indicate where $Z=\pm 1 H$ and $\pm 2H$. The black regions are where the dispersion relation does not have a physical solution, and thus the SWI is believed to be forbidden. The red dashed curves indicate where the local buoyancy frequency equals to a half of the doppler-shifted frequency of the spiral waves: $N^2 = (\omega_s/2)^2$. The upper half of each panel presents the wavelengths of unstable inertial modes in response to monochromatic $m=2$ waves, whereas the lower half of each panel shows those in response to monochromatic $m=3$ waves. We use $n=4$ as a representative example purely for the purposes of illustration, but the numbers on the colorbars can be linearly scaled to other modes as noted in the labels.}
\label{fig:dr}
\end{figure}

In Figure \ref{fig:dr}, we present two-dimensional maps showing the vertical and radial wavelengths of the inertial modes that satisfy the resonance conditions for monochromatic $m=2$ and $m=3$ spiral waves, assuming an adiabatic response of gas with $\gamma=1.4$ and the disk model that will be introduced in Section \ref{sec:model}.  
The wavelengths are calculated using Equations (\ref{eqn:k-spiral}), (\ref{eqn:kRi}), and (\ref{eqn:kZi2}).

We point out two important features from the figure.
First, as we have shown above, there is a limited region where the SWI can be triggered for a given monochromatic perturbation.
At the midplane, where the \brunt frequency becomes zero, the SWI can operate in the entire region inside of the ILR for $m=2$ waves ($R \lesssim 0.63~R_p$), whereas it can only operate in the region $0.48~R_p \lesssim R \lesssim 0.76~R_p$ for $m=3$ waves.
Second, when buoyancy is considered, the SWI-permitted regions are confined towards the midplane at the radii where the SWI can operate because the SWI can operate only in disk regions where $(\omega_s/2)^2 \lesssim N^2$ is satisfied \citep{bae16}.
The vertical extent of this SWI-permitted region is $\sim 1H$ above and below the midplane, and this is where we will focus on in this paper.
While $m \geq 3$ modes can trigger the SWI near the surface region ($|Z| \gtrsim 2H$) at $R \leq ((m-2)/m)^{2/3} R_p$ because of buoyancy, and this may have important implications for the morphology of spiral arms traced by observations of infrared scattered light, we do not consider this further here as we are mainly concerned with the mixing of solid particles located near the midplane.

\subsection{Planet-induced spiral waves}
\label{sec:planet-spirals}

A Jovian planet on a circular orbit will excite a spiral wave pattern that is a superposition of components with different azimuthal mode numbers, but with each component displaying a pattern speed, $\Omega_{\rm p}$, that is equal to the Keplerian angular velocity of the planet.
Nonlinear effects may cause the relation between the different components to vary as a function of radius. 
As shown later in the paper via a Fourier analysis, the spiral wave pattern at any radius interior to the Lindblad resonances due to a Jovian planet can be decomposed into a sum of modes where the dominant contributors are the $m=2$ and $m=3$ modes. 
At any point in time, one can assume that a protoplanetary disk will support the whole spectrum of possible inertial modes, which, in the absence of a strong excitation mechanism, will be present with very low amplitudes. 
The presence of a spiral wave pattern consisting of different azimuthal components, each with different Doppler-shifted frequencies as observed by fluid elements orbiting in the disk, will lead to excitation of those inertial modes that can resonantly interact with any of the azimuthal mode components that make up the incoming spiral wave. 
As such, we expect that a giant planet-induced spiral wave will lead to the excitation of inertial modes in a manner that is similar to a superposition of the response to both the $m=2$ and $m=3$ spiral modes described above. 
For lower mass planets, where the strengths of the spiral wave components will not be so strongly concentrated towards the $m=2$ and $m=3$ modes, the SWI may also be excited by the higher-$m$ modes.

Although our focus in this paper is on planets with circular orbits, it is worthwhile briefly discussing how the above picture changes if orbital eccentricity is considered. 
If the planet is on an eccentric Keplerian orbit, then $m$-fold spiral waves are excited at Lindblad resonances with pattern speeds $\Omega_{\rm p} = \left(m \pm k \right) \Omega_{\rm pl}/m$, where $k$ is a positive integer and $\Omega_{\rm pl}$ is the mean motion of the planet \citep{goldreich80}. 
When considering planet-disk interactions at Lindblad resonances, the strength of the torque exerted on the disk by the planet, $\Gamma$, scales as $\Gamma \sim e^2k$, where $e$ is the orbital eccentricity.
Therefore, increasing the planet eccentricity gives rise to new spiral waves, associated with higher values of $k$, being excited, with their amplitudes increasing as the eccentricity increases. 
Considering Lindblad resonances located interior to the planet that launch inward propagating spiral waves, these additional spiral waves can have higher pattern speeds than the equivalent $m$-fold spirals associated with planets on circular orbits. 
For example, taking $m=2$ and $k=1$, we have a wave that is excited with pattern speed $\Omega_{\rm p} = 3/2 \Omega_{\rm pl}$ instead of $\Omega_{\rm p}=\Omega_{\rm pl}$ as is the case for a planet on a circular orbit. 
Clearly, introducing the possibility of eccentric orbits increases the range of inertial mode frequencies that can be resonantly excited.

%figure 2
\begin{figure*}
\centering
\epsscale{1.15}
\plotone{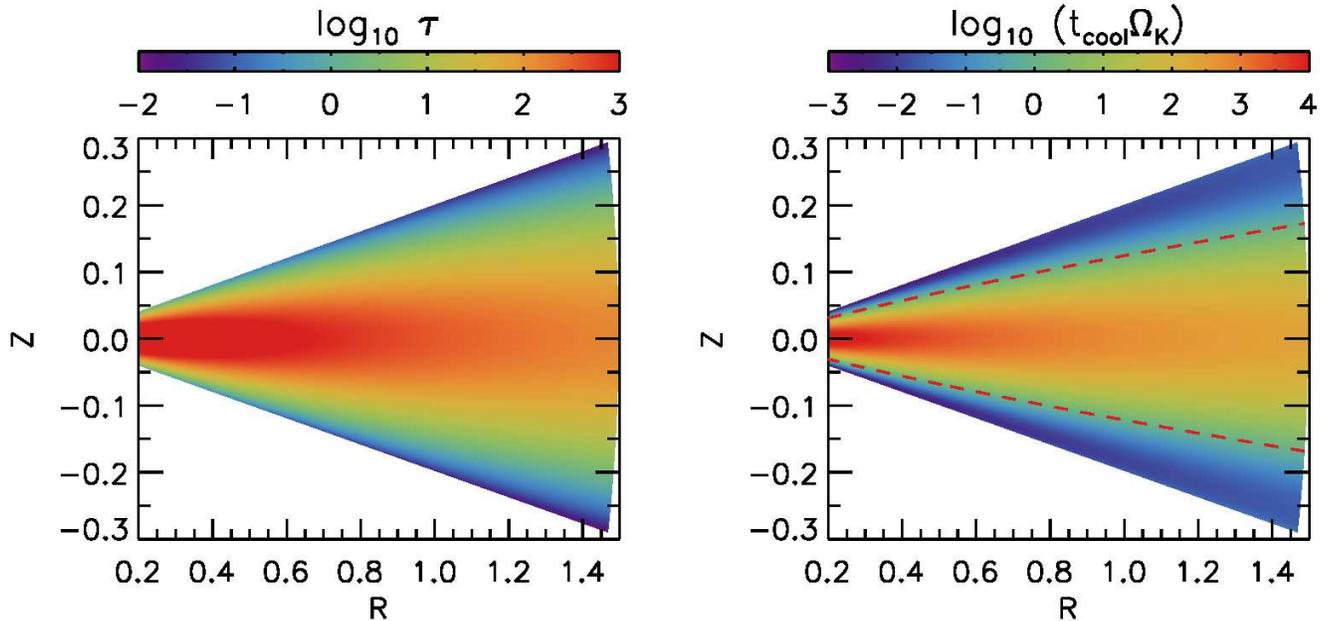}
\caption{Two-dimensional $R-Z$ distributions of (left) optical depth $\tau$ and (right) the dimensionless cooling time $\beta = t_{\rm cool}\Omega_K$ for the initial disk. The red dashed curves in the right panel indicate where $t_{\rm cool}\Omega_K = 1$. The main body of the disk is expected to behave adiabatically, since the cooling timescale is much longer than the dynamical timescale in the region.}
\label{fig:model}
\end{figure*}

\section{NUMERICAL METHODS}
\label{sec:method}

\subsection{Basic Equations}
\label{sec:eqns}

We solve the hydrodynamic equations for mass, momentum, and internal energy conservation in the three-dimensional spherical coordinates $(r,\theta,\phi)$:
\be\label{eqn:mass}
{\partial \rho \over \partial t} + \nabla \cdot (\rho v) = 0,
\en
\be\label{eqn:momentum}
\rho \left( {\partial v \over \partial t} + v \cdot \nabla v \right) = - \nabla P - \rho \nabla (\Phi_* + \Phi_p),
\en
\be\label{eqn:energy}
{\partial e \over \partial t} + \nabla \cdot (ev) = - P \nabla \cdot v + Q_{\rm cool}.
\en
In the above equations, $\rho$ is the mass density, $v$ is the velocity, $P$ is the pressure, $\Phi_* = GM_*/r$ is the gravitational potential from the central star, $\Phi_p$ is the external potential (see below), $e$ is the internal energy per unit volume, and $Q_{\rm cool}$ is the cooling rate (see below).

In the case of monochromatic spiral waves, we implement the following potential form:
\be
\label{eqn:potmodel1}
\Phi_p = \mathcal{A} \cos[m(\phi - \Omega_p t)] e^{-(R-R_p)^2/\sigma_p^2}.
\en
Here, $\mathcal{A}$ determines the spiral wave amplitude which is assumed to be constant over time $t$, $m$ is the azimuthal mode number, $\Omega_p$ is the pattern speed, $R=r\sin\theta$ is the cylindrical radius, $R_p$ is the radius about which the potential is centered, and $\sigma_p$ is the radial width of the potential.
We assume that the pattern speed is the local Keplerian frequency at its central position $\Omega_p = (GM_*/R_p^3)^{1/2}$ and that $\sigma_p = 0.2R_p$.

When a planetary companion is considered, its potential is included as 
\be
\label{eqn:potmodel2}
\Phi_p = -{{GM_p} \over {(|{\bf{r}}-{\bf{r_p}}|^2 + b^2)^{1/2}}},
\en
where $M_p$ is the planetary mass, ${\bf r}$ and ${\bf r_p}$ are the radius vectors of the center of grid cells in question and of the planet, and $b$ is the smoothing length.
We increase the planetary mass from zero to its full mass (i.e., $1~M_J$) over ten orbital times.
In three-dimensional calculations, the smoothing length is used only to avoid singularities in the potential on the grid scale. 
We thus adopt the cell diagonal size $((\Delta r)^2 + (r\Delta\theta)^2 + (r\Delta\phi)^2 )^{1/2}$ at the position of the planet as the smoothing length.
With the disk model introduced in Section \ref{sec:model}, the smoothing length corresponds to about $18 \%$ of a scale height, or about $13 \%$ of the Hill radius when the planet is fully grown to $1~M_J$.
In the planet run, we also include the indirect potential that arises because the origin of the coordinate system is based on the central star and not the center of the mass of the system.

We make use of an adiabatic equation of state.
The gas pressure and the internal energy are thus related through $P=(\gamma-1)e$, with an adiabatic index $\gamma=1.4$ adopted.
To realize the radiative cooling of the disk we implement a simple, but physically motivated, cooling scheme.
This assumes relaxation of the internal energy towards the background disk temperature $T_0$ at each location on the cooling timescale $t_{\rm cool}$.
Then, the cooling rate can be written as 
\be
\label{eqn:cooling1}
Q_{\rm cool} = - \rho c_v  {{T - T_0} \over {t_{\rm cool}}},
\en
where $c_v$ denotes the heat capacity at constant volume.

The cooling timescale is calculated every time step for each grid cell, using the optical depth and the temperature.
We follow the approach used in \citet{lyra16} and previously in \citet{lyra10} and \citet{horn12} in a vertically-integrated two-dimensional approximation.
To briefly summarize, the cooling timescale is estimated to be the radiative timescale that is defined as 
\be
\label{eqn:coolingtime}
t_{\rm cool} \equiv {{\int e dV} \over {\int F \hat{{\bf n}} \cdot d{\bf A}}},
\en
where $e = \rho c_v T$ and $F = |{\bf F}| = \sigma T^4 / \tau_{\rm eff}$ with $\sigma$ and $\tau_{\rm eff}$ being the Stefan-Boltzmann constant and the effective optical depth.
The effective optical depth $\tau_{\rm eff}$ is given by 
\be
\label{eqn:tau}
\tau_{\rm eff} = {3 \over 8} \tau + {\sqrt{3} \over 4} + {1 \over 4\tau}
\en
to take account of both optically thick and thin limits \citep{hubeny90,dangelo03}.

The integration in Equation (\ref{eqn:coolingtime}) is done over a sphere that has a radius equal to the local pressure scale height.
This results in a cooling time of 
\be
\label{eqn:coolingtime2}
t_{\rm cool} = {{\rho c_v H \tau_{\rm eff}} \over {3\sigma T^3}}.
\en

To obtain the optical depth $\tau$, we first calculate the optical depth due to the material above and below each grid cell as
\be
\tau_{\rm upper} = \int_{z}^{z_{\rm max}} {\rho(z') \kappa(z')} dz'
\en
and
\be
\tau_{\rm lower} = \int_{z_{\rm min}}^{z} {\rho(z') \kappa(z')} dz'.
\en
In practice, the integration is done in $\theta$, instead of $z$, for simplicity.
Then, we calculate the optical depth as $1/\tau = 1/\tau_{\rm upper} + 1/\tau_{\rm lower}$ to give a correct midplane cooling rate \citep{lyra16}.
For the opacity $\kappa$ in the above equations, we use the Rosseland mean opacity of \citet{zhu09}.

The optical depth and the cooling time of the initial disk are presented in Figure \ref{fig:model}.
As shown, the main body of the disk is optically thick, with the cooling time much longer than the dynamical time: $t_{\rm cool} \gg 1/\Omega_K$.
Therefore, we expect that the main body of the disk behaves essentially adiabatically, and hence will be stable against the growth of the vertical shear instability \citep{nelson13}.

Our calculations are inviscid, but artificial viscosity and the associated heating are included in the momentum and internal energy equations \citep{stone92}. 
As discussed in \citet{bae16}, the effective dimensionless kinematic viscosity associated with the numerical diffusion for FARGO3D \citep{benitez16} operates with a value that is below $\nu=10^{-6}$.

\subsection{Disk Models}
\label{sec:model}

We begin with an initial radial power-law temperature distribution in the disk that is independent of height:
\be
\label{eqn:temp}
T(R) = T_p \left( {R \over R_p} \right)^q,
\en 
where $T_p$ is the temperature at the location of the perturber $R=R_p$.
The isothermal sound speed is related to the temperature by $c_s^2=\mathcal{R} T/\mu$, where $\mathcal{R}$ is the gas constant and $\mu$ is the mean molecular weight.
Thus, Equation (\ref{eqn:temp}) corresponds to the radial sound speed distribution given by
\be
c_s(R) = {H_p \over R_p} \left( {R \over R_p} \right)^{q/2}.
\en 
In all models, we adopt $H_p/R_p = 0.05$ and $q=-1$ such that the entire disk has the aspect ratio of 0.05.

The initial density and azimuthal velocity profiles are constructed to satisfy hydrostatic equilibrium \citep[e.g.,][]{nelson13}: 
\be
\label{eqn:init_den}
\rho(R,Z)   =  \rho_p \left({R \over R_p}\right)^{p} \exp\left({GM_* \over c_s^2} \left[{1 \over \sqrt{R^2+Z^2}} - {1 \over R}\right]\right)
\en
and
\be
\label{eqn:init_vel}
v_\phi(R,Z)  =\left[ (1+q){GM_* \over R} + \left( p + q \right) {c_s^2} - q{GM_* \over \sqrt{R^2+Z^2}} \right] ^{1/2}.
\en
We choose the initial density distribution in such a way that the vertically-integrated surface density $\Sigma$ becomes $\Sigma = 5 \Sigma_{\rm MMSN}$, where $\Sigma_{\rm MMSN}$ is the minimum mass solar nebular (MMSN) model of \citet{hayashi81}. 
A value of $p=-5/2$ is adopted accordingly, to have the surface density power-law slope that matches to the slope of the MMSN model, $-3/2$.

The initial radial and meridional velocities are set to zero, but uniformly distributed random perturbations are added as white noise, at the level of $10^{-6}~c_s$, to the initial meridional velocity field in order to seed the instability.

\subsection{Computational Setup}
\label{sec:setup}

We expect the main body of the disk to behave adiabatically, since the disk cooling timescale is much longer than the dynamical timescale in the region (see Figure \ref{fig:model}).
Thus, the regions in which the SWI is allowed to operate can be inferred using the linear analysis described in Section \ref{sec:expectation} and depicted in Figure \ref{fig:dr}, assuming an adiabatic equation of state with $\gamma=1.4$.

As seen in Figure \ref{fig:dr}, the SWI-permitted regions are confined around the midplane, with a thickness that is as small as $\sim 1H$ above and below the midplane.
Therefore, we aim to have a numerical resolution with which unstable inertial modes with $\lambda_Z \simeq 1H$ can be well captured.
At the midplane, the ratio of the vertical to radial wavelengths of unstable inertial modes increases over radius, except near the ILR where a singularity appears.
The largest ratio of $\lambda_Z / \lambda_R$ for perturbations with $m \geq 2$ is $\simeq 1.5$ (see Figure \ref{fig:dr}), and thus the radial wavelength that we aim to resolve will be $\lambda_R \simeq (2/3)H$.

Since inertial modes can have arbitrarily small length scales, smaller scale unstable modes can only be captured with higher resolutions than the one used in the present work. On the other hand, it seems that the modes with the largest length scale contain the largest amount of kinetic energy \citep{bae16}.
In the simulations presented in Section \ref{sec:results} as well as the ones in \citet{bae16}, we observe that smaller scale modes grow at earlier times, but only with small velocity amplitudes.
These modes are then masked by larger scale modes that grow at later times with larger velocity amplitudes, supporting the conjecture that the modes with the largest length scales contain most of the kinetic energy as the system approaches the nonlinear saturated state. 
One effect of under-resolving the small scale modes, however, is that the development of a realistic breakdown into hydrodynamic turbulence may be prevented, as the cascade of energy from large to small scales in the fully saturated nonlinear regime will be prevented, affecting the nature of the flow and the statistics of the quasi-turbulent flow that develops due to the SWI.

Our simulation domain extends from $r_{\rm in} = 0.2 R_p$ to $r_{\rm out} = 1.5 R_p$ in radius, from $\pi/2-0.2$ to $\pi/2+0.2$ in the meridional direction (covering 4 scale heights above and below the midplane), and from 0 to $2\pi$ in azimuth.
We have tested with various numerical resolutions and find that 12 grid cells per wavelength is required to properly resolve unstable inertial modes.
To achieve $\Delta r = (1/12) \lambda_R = (1/12) \times (2/3) H$, or equivalently $\Delta r / r = (1/18) H/R$ at the midplane, we adopt 726 logarithmically-spaced radial grid cells.
We adopt 144 and 754 uniformly-spaced grid cells in the meridional and the azimuthal directions, respectively, with which choice $\Delta r : r \Delta\theta : r \Delta\phi \simeq 1:1:3$.

%figure 3
\begin{figure*}
\centering
\epsscale{1.1}
\plotone{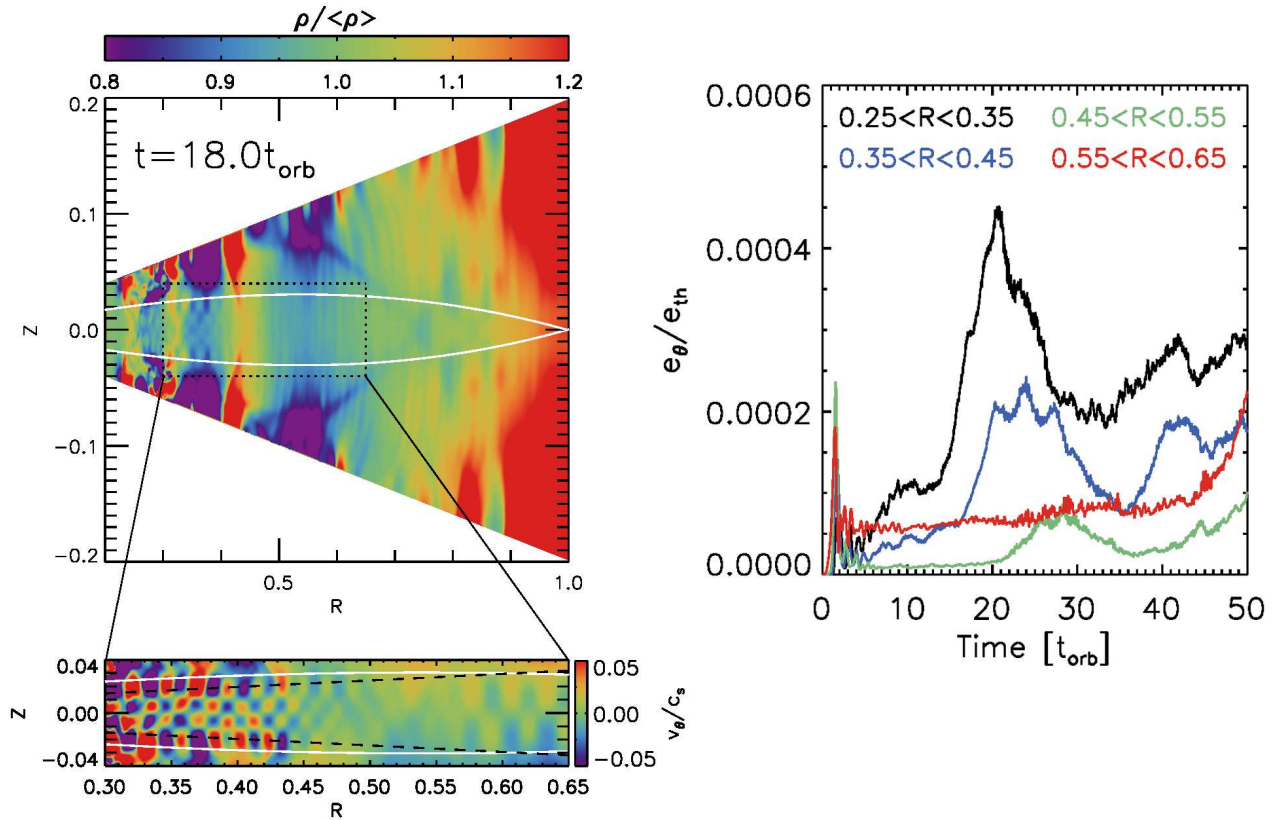}
\caption{(Left Top) Two-dimensional distribution of density perturbation $\rho / \langle \rho \rangle$ in a $R-Z$ plane for $m=2$ model. The two white curves near the midplane indicate where the local \brunt frequency equals to one half of the doppler-shifted wave frequency of spiral waves  ($N^2 = (\omega_s/2)^2$), between which region is expected to be unstable to the SWI. (Left Bottom) Contour plots of the meridional velocity normalized by the local sound speed, for the dotted rectangle region in the left top panel. The $x$ and $y$ axes are drawn isotropically so that the radial and vertical wavelengths of unstable inertial modes can be compared with each other from the figure. The black dashed lines indicate where $Z = \pm 1 H$. Note that the SWI has grown in the entire disk region inside the ILR. (Right) Time evolution of the meridional kinetic energy $e_\theta$ at various radius bins, normalized by $e_{\rm th}$ in each bin. Note that the exponential growth of the instability appears at earlier times for smaller radii.}
\label{fig:pot2}
\end{figure*}

The radial boundary condition is chosen to have a zero gradient for all variables.
We further implement a wave damping zone \citep{devalborro06} in the intervals $r=[r_{\rm in}, 1.2r_{\rm in}]$ and $r=[0.9r_{\rm out},r_{\rm out}]$, since reflected waves at the radial boundaries may affect triggering and growth of the SWI, as they have the same pattern speed as the incoming waves.
Periodic boundary conditions are used in azimuth since the simulation domain covers $2\pi$.
At the meridional boundaries, we use an outflow boundary condition such that all velocity components in the ghost zones have the same values as the last active zones, but the meridional velocity is set to 0 if directed toward the disk midplane. 
The temperature in the ghost zones is set to have the same value as in the last active zones. 
The density in the ghost zone is then obtained by solving the hydrostatic equilibrium in the meridional direction:
\be
\label{eqn:hsebc}
{1 \over \rho} {\partial \over \partial\theta}{(\rho c_s^2)} = {v_\phi^2 \over \tan\theta}.
\en 

We make use of FARGO3D \citep{benitez16}, which runs on both central processing units (CPUs) and graphics processing units (GPUs).
In order to deal with the high numerical resolution, our calculations use a cluster of GPUs, which allows us to perform the calculations within a reasonable time frame thanks to a large speed up with respect to CPUs.
With four NVIDIA Tesla K20x GPUs on the University of Michigan high-performance computing cluster\footnote{http://arc-ts.umich.edu/systems-and-services/flux/}, it takes about one month for the planet run in Section \ref{sec:planet} to evolve for 200 orbits. 
We enable the FARGO (Fast Advection in Rotating Gaseous Objects) orbital advection module \citep{masset00}.

In the following sections we will use the orbital distance of the planet (or perturbing potential) $R_p$ as the length unit (assumed to be 5~au), and the orbital time $1~t_{\rm orb}$ at $R=R_p$ as the time unit.

Since we adopt physical units to consider cooling, the models are not scalable in principle.
However, we believe that the results can be qualitatively applicable as far as the main body of a disk is sufficiently optically thick, because the disk response to spiral waves is not very sensitive to the cooling timescale when $\beta \equiv t_{\rm cool}\Omega_K \gtrsim 1$ \citep[][and also from our test runs during early phases of the present work]{zhu15}.

%figure 4
\begin{figure*}
\centering
\epsscale{1.1}
\plotone{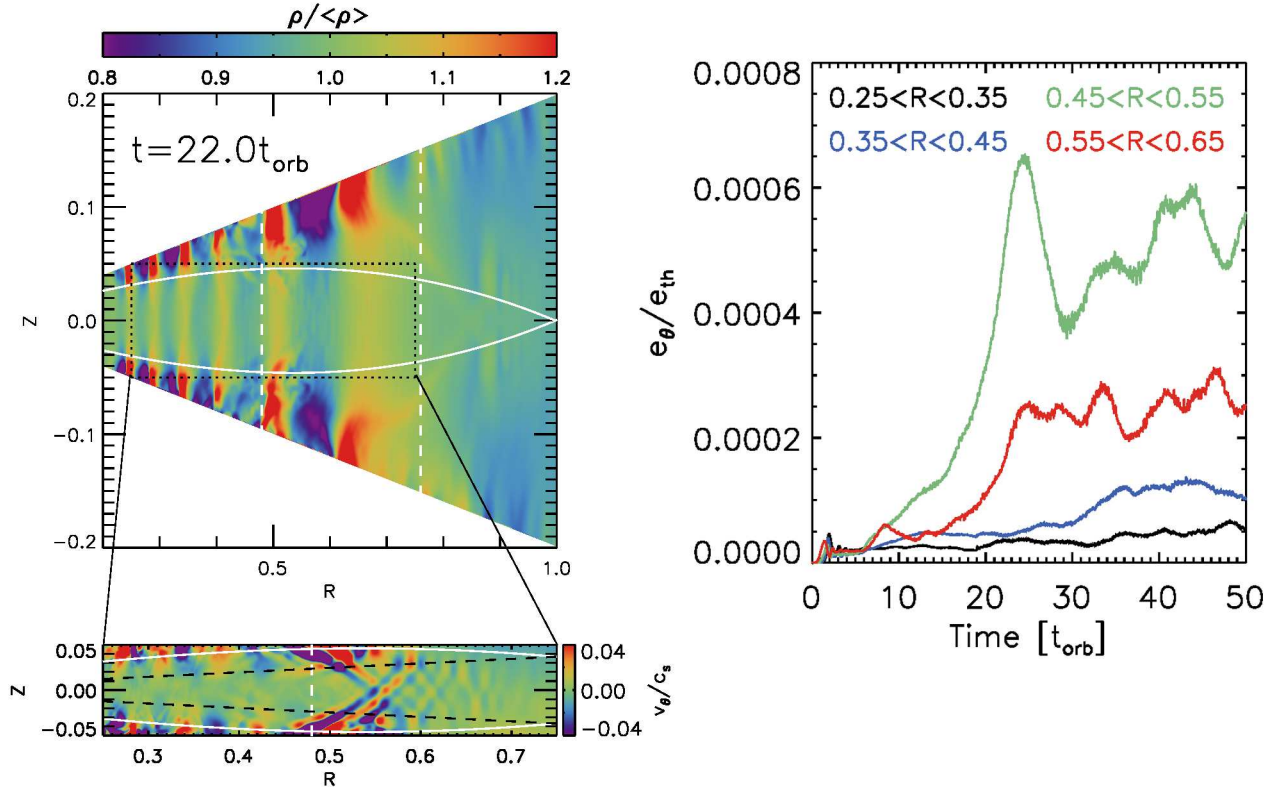}
\caption{Same as Figure \ref{fig:pot2}, but for $m=3$ model. Note that the instability has grown beyond $R \sim 0.5$, but not inward of the radius as no inertial modes there have an adequate frequency to satisfy the resonant condition with the incoming spiral waves. Also, the exponential growth in $e_\theta$ appears for $0.45 < R < 0.55$ and $0.55 < R < 0.65$ but not at the inner two radius bins, supporting the fact that the SWI does not operate at $R \lesssim 0.5$. The increase in $e_\theta$ in the inner two radius bins at later times ($t \gtrsim 30~t_{\rm orb}$) is presumably because the spiral waves propagate through the SWI-turbulent region.}
\label{fig:pot3}
\end{figure*}

\subsection{Diagnostics}

In order to examine the growth and the saturation of the SWI, we compute the volume-integrated meridional kinetic energy $e_\theta$: 
\be
e_{\theta} = {1 \over 2} \int_{V}{\rho v_{\theta}^2}~dV.
\en
While the velocity perturbation driven by spiral waves is larger at high altitude than near the midplane in general, $e_\theta$ traces perturbations near the midplane region because of the vertically stratified density structure.

In the figures presenting the meridional kinetic energy, we normalize $e_\theta$ by $e_{\rm th}$ which we define as the volume-integrated kinetic energy of gas parcels assuming that they move at the local sonic velocity:
\be
e_{\rm th} = {1 \over 2} \int_{V}{\rho {c_s}^2}~dV.
\en

We also compute the Shakura-Sunyaev stress parameter $\alpha_{r \phi}$ as 
\be
\alpha_{r\phi}(r, \theta) = { {{\langle \rho \delta v_r \delta v_\phi \rangle} \over  \overline{P}(r) }},
\en
where $\overline{P}(r)$ is a density-weighted mean pressure at radius $r$, in order to examine the sustained level of 
angular momentum transport driven by the turbulent flow at saturation of the instability (although we note that a contribution to $\alpha_{r\phi}$ also arises from the propagating spiral waves that is difficult to disentangle from that arising from the SWI-induced turbulence).

\section{RESULTS}
\label{sec:results}

\subsection{Results with Monochromatic Waves}

Before we describe our main results with a planet, we introduce models with monochromatic perturbations.
The perturbing waves are imposed using Equation (\ref{eqn:potmodel1}) with azimuthal mode numbers $m=2$ (Section \ref{sec:pot2}; hereafter $m=2$ model) or $m=3$ (Section \ref{sec:pot3}; hereafter $m=3$ model).
The potential amplitudes are chosen so that the linear phase of the instability spans several orbital times, and therefore the growth of individual unstable modes can be captured: $\mathcal{A} = 4 \times10^{-3}$ for the $m=2$ model and $\mathcal{A} = 5 \times 10^{-4}$ for the $m=3$ model.

\subsubsection{$m=2$ Model}
\label{sec:pot2}
In Figure \ref{fig:pot2}, we present the density distribution $\rho / \langle \rho \rangle$ for $m=2$ model, where the brackets denote an azimuthal average, along with the meridional velocity distribution near the midplane where the SWI is expected to operate.
The density perturbation is about $10 \%$ in the midplane and about $30 \%$ at the surface when the spiral waves are fully established.
The larger perturbation towards the surface is presumably because of a nonlinear effect due to faster advection of the wave at higher altitudes in the disk \citep[][see also Zhu et al. 2015]{bae16}.

The checkerboard pattern shown in the meridional velocity distribution is a generic feature of the linear growth phase of the SWI \citep{bae16}, and was also observed by \cite{fromang07} in their shearing box simulations of nonlinear, axisymmetric sound waves propagating in astrophysical disks. 
As expected from the dispersion relations, the regions in which the SWI develops are confined toward the midplane, because the large \brunt frequency near the surface does not allow any inertial modes there to resonantly interacting with the incoming $m=2$ waves.
During the linear phase, the magnitude of the perturbed meridional velocity increases over time, but remains a few to about ten percent of the local sound speed.
The wavelengths of the dominant unstable inertial modes varies over radius: at the midplane, $\lambda_R \simeq \lambda_Z \simeq 1 H$ at $R \sim 0.4$, whereas $\lambda_R \simeq 1H$ and $\lambda_Z \simeq 2H$ at $R \sim 0.6$.
The ratio of vertical to radial wavelengths of the unstable inertial modes increases toward larger radii and height. 
Note that these unstable mode properties are well described by the linear analysis introduced in Section \ref{sec:expectation}.

Also shown in Figure \ref{fig:pot2} is the time evolution of the volume-integrated meridional kinetic energy $e_\theta$ for various radius bins.
The exponential growth of $e_\theta$, for example during $t \sim 15-21~t_{\rm orb}$ at $0.25<R<0.35$, indicates the development of the SWI. 
At smaller radii the instability becomes apparent at earlier times because of the shorter interaction period between inertial modes and the spiral waves.

\subsubsection{$m=3$ Model}
\label{sec:pot3}

Figure \ref{fig:pot3} displays the density distribution $\rho / \langle \rho \rangle$ for the $m=3$ model.
The density perturbation is about $5 \%$ in the midplane and about $20 \%$ at the surface when the $m=3$ spiral waves are fully established.
As seen in the figure, the instability is developing beyond $R \gtrsim 0.5$ at $t=22~t_{\rm orb}$. 
The SWI does not operate inward of this radius because the Doppler-shifted perturbation frequency from the incoming $m=3$ waves is too large to excite any inertial modes, which is in very good agreement with the expectation from the dispersion relations.
As in the $m=2$ model, the ratio of vertical to radial wavelength of unstable inertial modes increases towards larger radii and height.

The time evolution of $e_\theta$ is presented in Figure \ref{fig:pot3}.
The exponential growth of $e_\theta$ is only observed in $0.45<R<0.55$ and $0.55 < R < 0.65$, supporting the conjecture that the instability is forbidden at smaller radii.
The increase in $e_\theta$ at later times (e.g., $t \gtrsim 30~t_{\rm orb}$) in the inner disk regions of $R < 0.45$ is probably because the spiral waves propagate through the turbulent region of $R > 0.45$, and hence generate some vertical motion there due to the curvature of the spiral wave fronts in the meridional plane discussed in \cite{bae16}.

\subsection{Results with a Perturbing Planet}
\label{sec:planet}

We now describe the results for the planet run.
We first discuss the overall evolution in Section \ref{sec:overall} and then focus on the development of the SWI in Section \ref{sec:pswi}.

%figure 5
\begin{figure}
\centering
\epsscale{1.15}
\plotone{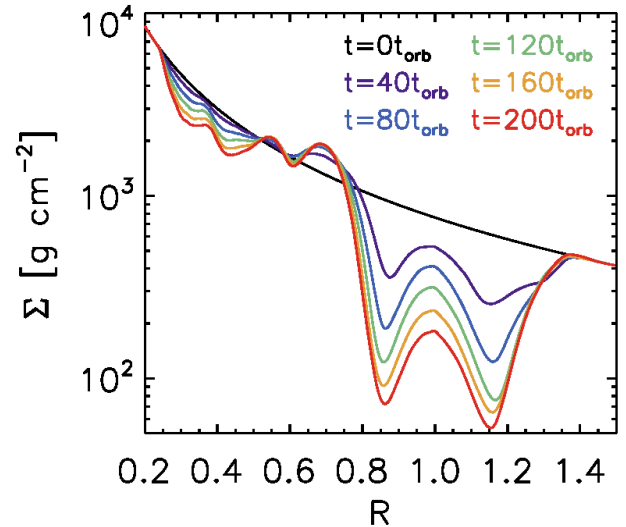}
\caption{Radial distributions of the azimuthally averaged surface density at $t=0$, 40, 80, 120, 160, and $200~t_{\rm orb}$.}
\label{fig:sigma}
\end{figure}

%figure 6
\begin{figure}
\centering
\epsscale{1.15}
\plotone{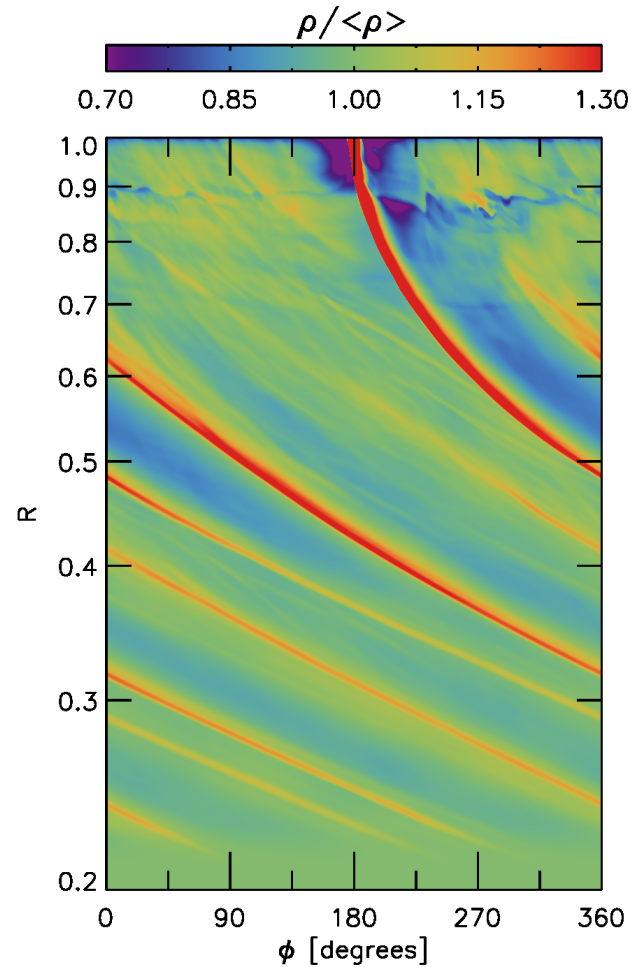}
\caption{Two-dimensional $\phi-R$ distribution of $\rho / \langle \rho \rangle$ in the midplane, taken when the planetary mass is fully grown to $1~M_J$ at $t=10t_{\rm orb}$. The planet is located at $(\phi, R) = (180, 1.0)$. The $y$-axis is plotted using a logarithmic scale to stretch the inner disk.}
\label{fig:relden}
\end{figure}

\subsubsection{Overall Evolution}
\label{sec:overall}

We evolve the planet run for $200~t_{\rm orb}$.
In Figure \ref{fig:sigma}, we present the azimuthally averaged surface density distributions at every $40~t_{\rm orb}$.
The planet opens a gap around its orbit and creates a mild density bump at the inner gap edge.
In the inner disk ($R \lesssim 0.5$), the disk looses material through the meridional boundary because of vertical flows induced by the spiral waves. 

As inferred from the figure, the depth of the gap does not reach full saturation by the end of the run at $200~t_{\rm orb}$.
Recently, \citet{fung16} carried out three-dimensional calculations to examine gap opening by planets in a locally isothermal disk.
Their results indicate that the saturation of the gap depth requires about 1000 planetary orbits (or even longer depending on the planetary mass) with a disk viscosity of $\alpha=10^{-3}$.
On the other hand, in their $1~M_J$  case, the density perturbation driven by spiral waves remains roughly constant after 100 planetary orbits (Fung, D., private communication). It is therefore reasonable to assume that the amplitudes of the spiral waves in our simulations have also saturated.

In Figure \ref{fig:relden}, we present the density distribution $\rho / \langle \rho \rangle$ in the midplane at $t=10~t_{\rm orb}$, at which time the planetary mass is fully grown to $1~M_J$.
The planet excites three distinguishable spiral arms. 
This multi-armed feature can be understood as a consequence of non-linear mode-coupling \citep[e.g.,][]{artymowicz92,lee16} and has been shown to arise in recent numerical simulations \citep[e.g.,][]{fung15,juhasz15,zhu15}.
Looking in more details at the spiral arms, the first arm is connected to the planet, and the second and third arms start to appear at $R \sim 0.7$ and at $R \sim 0.6$, respectively.
The azimuthal separation between the spiral arms, as well as the relative strength of the arms, vary over radius.
At $R = 0.4$, for instance, the second arm is $\sim100^\circ$ ahead of the first arm and the third arm is $\sim100^\circ$ behind the first arm.
The density perturbation by the second arm at this radius is about twice as strong as the perturbation driven by the other two arm.
On the other hand, at $R = 0.3$, the second arm is $\sim75^\circ$ ahead of the first arm and the third arm is $\sim120^\circ$ behind the first arm.
The density perturbation by the second arm is still the strongest, but only $\sim25\%$ stronger than the perturbation driven by the first arm.

%figure 7
\begin{figure}
\centering
\epsscale{1.15}
\plotone{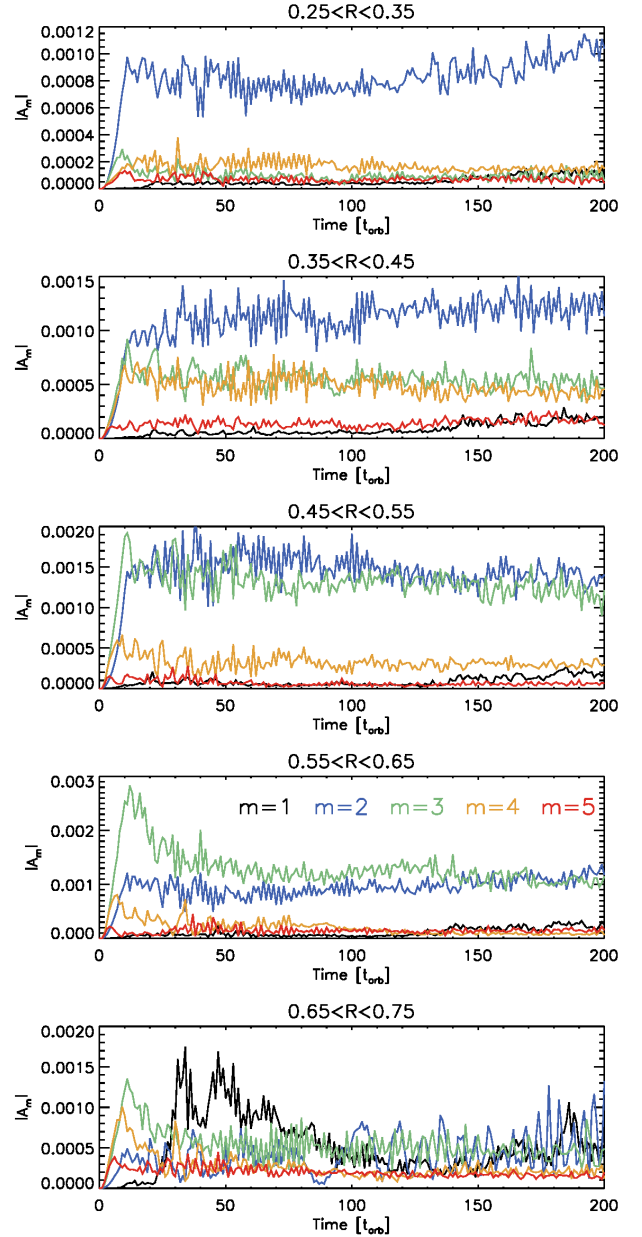}
\caption{Time evolution of the Fourier amplitudes $|A_m|$ for various azimuthal modes ($m = 1-5$) in different radius bins. Note that the amplitudes remain nearly constant after $\sim100~t_{\rm orb}$.}
\label{fig:fpower}
\end{figure}

In order to more quantitatively examine the gas response to the spiral waves, in Figure \ref{fig:fpower} we plot the time evolution of the Fourier amplitudes of the density perturbation at different radii.
As seen in the figure, spiral waves driven by a planet are not monochromatic, but consist of various azimuthal components that are superimposed on each other. 
We emphasize that, most importantly for our purpose, the Fourier amplitudes of $m=1-5$ modes remain nearly constant after $t \sim 100~t_{\rm orb}$ at all radii, despite the fact that the gap depth has increased by about a factor of two in between $100~t_{\rm orb}$ and $200~t_{\rm orb}$.
Another important feature is that the strongest mode shifts toward smaller azimuthal wave numbers at smaller radii.
The $m=3$ mode has the largest Fourier amplitude at $0.55 < R <0.65$, the $m=2$ and $m=3$ modes have a comparable amplitude at $0.45 < R < 0.55$, and the $m=2$ dominates at $0.25 < R < 0.45$. For completeness, we note that there are no $m=1$ spiral modes that propagate interior to the planet because formally the $m=1$ ILR is not present in the disk. 
The appearance of a finite amplitude $m=1$ Fourier component arises presumably because the disk generates an $m=1$ perturbation due to the appearance of a low amplitude vortex, or because the disk becomes mildly eccentric.

%figure 8
\begin{figure*}
\centering
\epsscale{1.15}
\plotone{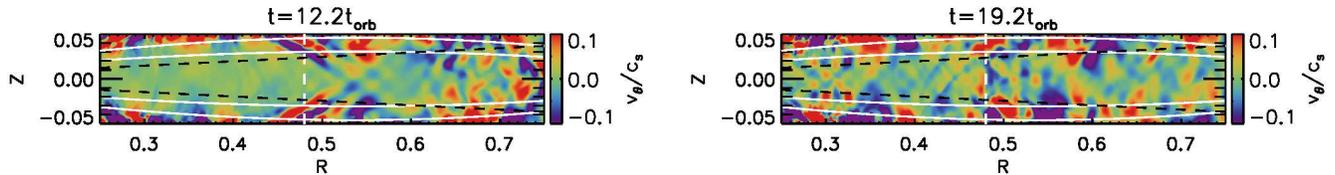}
\caption{Two-dimensional distributions of the meridional velocity in a $R-Z$ plane at (left) $t=12.2 ~t_{\rm orb}$ and (right) $t=19.2~t_{\rm orb}$. The two sets of white curves indicate where the local \brunt frequency equals to one half of the Doppler-shifted wave frequency, for monochromatic $m=2$ waves (the ones close to the midplane) and $m=3$ waves (the ones close to the surface). The vertical dashed line indicates the radial location inward of which the SWI is forbidden for monochromatic $m=3$ waves. The black dashed lines denotes where $Z = \pm 1H$. The two axes are drawn isotropically.}
\label{fig:vz_planet}
\end{figure*}

%figure 9
\begin{figure*}
\centering
\epsscale{1.15}
\plotone{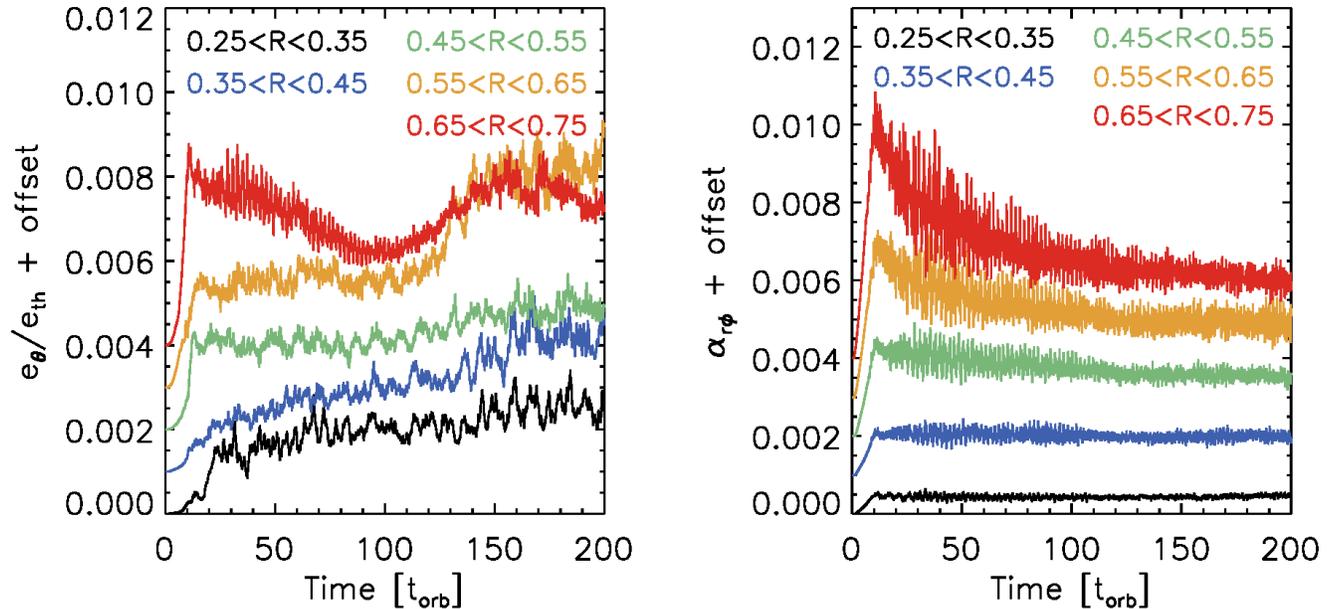}
\caption{(Left) Time evolution of the meridional kinetic energy $e_\theta$ at various radius bins, normalized by $e_{\rm th}$ in each bin. (Right) Time evolution of the Reynolds Stress $\alpha_{r\phi}$ at various radius bins. In both panels, we add vertical offsets of 0.001, 0.002, 0.003, and 0.004 to the blue, green, yellow, and red curves for illustration purposes.}
\label{fig:evol}
\end{figure*}

\subsubsection{Spiral Wave Instability}
\label{sec:pswi}

The SWI starts to grow from very early times even before the planetary mass is fully grown ($t \sim 4-5~t_{\rm orb}$) at $R \gtrsim 0.5$, because the perturbation is strong enough there even with a sub-Jovian mass planet.
The weak, wave-like inter-arm structures at $R \gtrsim 0.5$ and the break-up of the second arm at $R \sim 0.6-0.7$ seen in Figure \ref{fig:relden}, are some of the indications of the instability.

In Figure \ref{fig:vz_planet}, we display the meridional velocity field in a vertical plane at $t=12.2~t_{\rm orb}$ and $t=19.2~t_{\rm orb}$.
In the snapshot taken at $t=12.2~t_{\rm orb}$, one can see the checkerboard pattern suggesting that the SWI is growing there, although the individual unstable modes are not as clearly identifiable as in the monochromatic case due to the complex nature of planet-driven waves (i.e., superposition of various azimuthal modes). 
However, we note that there is no signature of the SWI in $R \lesssim 0.5$ at this time epoch.
This is probably because (1) the perturbation in the region is small compared with the $R \gtrsim 0.5$ region (Figure \ref{fig:relden}) and/or (2) the $m=3$ component, with which the SWI is forbidden in this region, has a comparable power to the $m=2$ mode during the early evolution (e.g., $0.35 < R < 0.45$ region in Figure \ref{fig:fpower}) so the growth rate of the instability could be reduced.
At $t=19.2~t_{\rm orb}$, checkerboard patterns appear in the inner disk region of $R \lesssim 0.5$.
The steep increase in the meridional kinetic energy around $t=20~t_{\rm orb}$, particularly for the $0.25 < R < 0.35$ region, presented in Figure \ref{fig:evol} supports the suggestion that the SWI is developing in the region.

%figure 10
\begin{figure*}
\centering
\epsscale{0.95}
\plotone{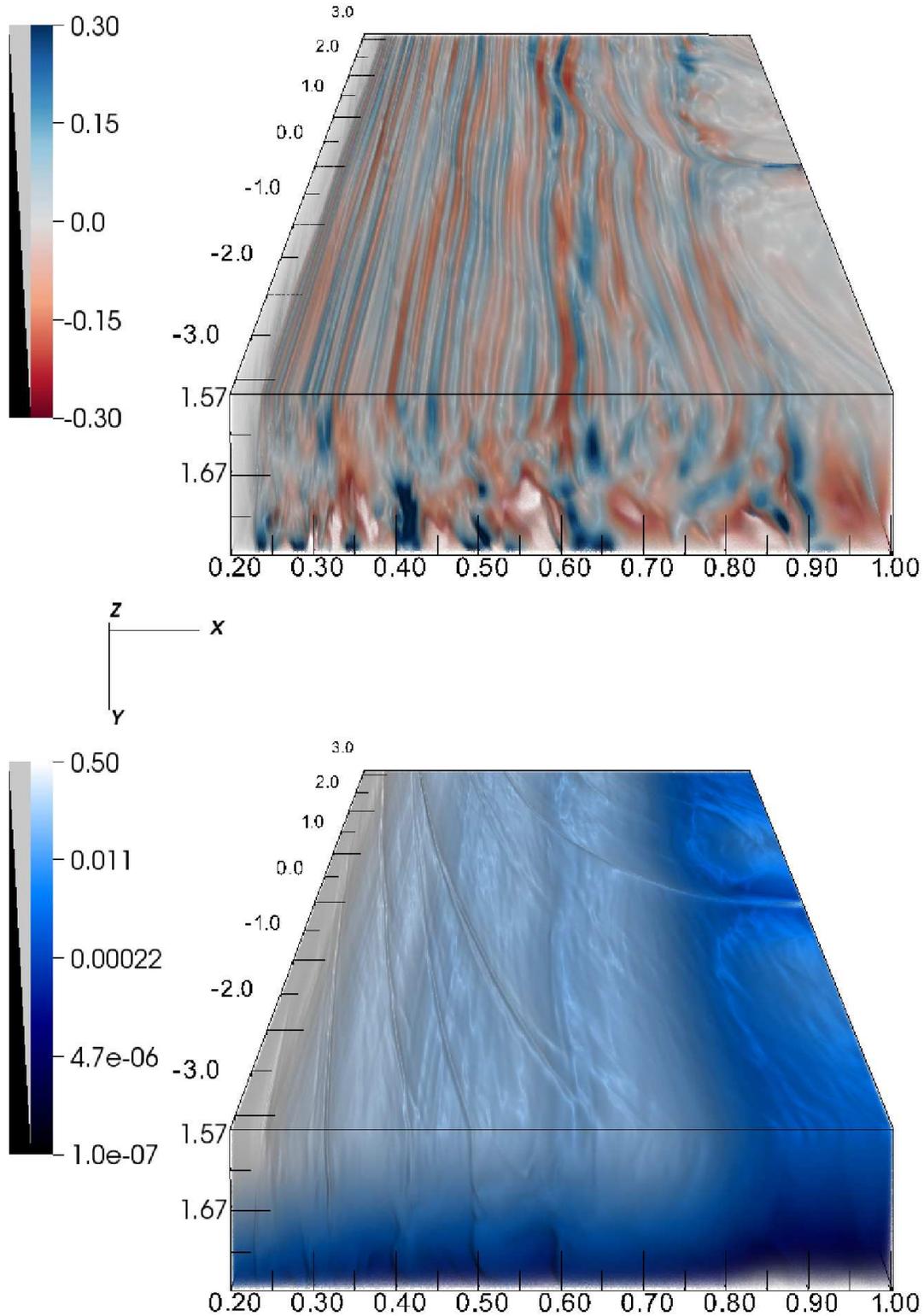}
\caption{(Top) Three-dimensional global view of the meridional velocity field in the bottom half of the simulations domain: the upper surface shows the disk midplane. The quantity is plotted in units of the local sound speed: $v_\theta/c_s$. $r$ on the X-axis, $\theta$ on the Y-axis in units of radians, and $\phi$ on the Z-axis in units of radians. Note that the instability creates radially-alternating vertical flows when saturated. The vertical flows have a magnitude of order of a few tens of percent at the midplane. (Bottom) A volume-rendering view of the density in a logarithmic scale. The snapshots are taken at the end of the simulation ($t=200~t_{\rm orb}$). The planet is located at $(X, Y, Z) = (1.0, \pi/2, 0.0)$.}
\label{fig:render}
\end{figure*}

As shown in Figure \ref{fig:evol}, the overall perturbed vertical kinetic energy and the stress associated with the combined action of the propagating spiral waves and the SWI (we have not sought to disentangle the contributions from these in this paper) reach quasi-steady state after $\sim 100~t_{\rm orb}$. 
This adds further evidence that the strength of the SWI is probably not very sensitive to the gap depth after this point as the spiral wave amplitudes have reached steady values.
We note that $e_\theta$ increases after $\sim120~t_{\rm orb}$ for $0.55 < R < 0.75$.
However, we do not find any significant changes in the strength of spiral arms or background disk structures to explain this rise.
One possible explanation is that some large-scale inertial modes, which have to have smaller growth rates than small-scale modes \citep{fromang07}, become unstable at this time, although we were not able to identify any corresponding individual mode growing as the disk has already developed strong vertical flows by this time, which we will discuss now.

In Figure \ref{fig:render}, we present a three-dimensional global view of the meridional velocity and the density at the end of the simulation ($t=200~t_{\rm orb}$).
As shown, the checkerboard meridional velocity patterns merge to create radially alternating vertical flows when the SWI saturates. 
Such vertical flows have been pointed out in the non-stratified, isothermal cylindrical models of \citet{bae16}, in which other sources capable of inducing vertical motion, such as the vertical shear instability that is known to cause the excitation of corrugation modes \citep{nelson13}, are absent.
We thus believe that the alternating vertical flow is a generic outcome of the SWI. Visual inspection of the upper panel of figure \ref{fig:render} suggests the azimuthal structure of these modes is of low degree, corresponding to perhaps a mixture of $m=0$ corrugation-type modes and/or $m=1$ warp or tilt modes, which are inertial modes for which the vertical velocity perturbations have no nodes in the vertical direction. In the case of $m=0$ corrugation modes, the whole disk column at a given radius, including the midplane, moves up and down in a coherent manner. For $m=1$ tilt modes the disk acts as a series of narrow rings each of which tilts rigidly. As these disturbances arise in the form of waves, they propagate through the disk as corrugation or warping disturbances. 
The vertical length scale of the vertical flows is similar to the thickness of the disk region in which the SWI is permitted, suggesting that it is set by the wavelength of the largest unstable inertial modes that are allowed to grow. 
So these might be modes which are excited directly by the SWI, or they may be the result of nonlinear mode coupling which allows energy to seep into these modes from other modes that are excited by the SWI. 
Being coherent and quite large scale, once excited these corrugation/tilt modes become a prominent feature of the flow because they do not damp efficiently. 
We find that the associated vertical flows can have perturbed vertical velocities that are in the range $\sim 30 - 50 \%$ of the local sound speed. 
One caveat, however, which we have already alluded to in Section~\ref{sec:setup} and which causes us to be cautious in interpreting our results, is that a resolution of $\sim 18$ cells per scale height in the vertical and radial directions does not allow for a turbulent cascade to develop efficiently. 
While some transfer of energy to small scales undoubtedly arises in our simulations, leading to dissipation on the grid scale, it may also be the case that the low resolution also favors the development of the SWI in such a manner that the large scale modes are more prominent than they would be in a more highly resolved simulation. 
Testing the outcome of the SWI as a function of resolution is a task that will be undertaken in the future when available computational resources allow such a study to be conducted.

\section{DISCUSSION}
\label{sec:discussion}

\subsection{Particle Stirring Induced by the SWI}
\label{sec:diffusion}

We measure the vertical diffusion coefficient to estimate the rate of vertical mixing of dust particles induced by the SWI.
In order to do this, we restart the planet run described in Section \ref{sec:planet} from $t=200t_{\rm orb}$, with outputs of the meridional velocity in three dimensions every $0.005~t_{\rm orb}$.
We calculate the vertical diffusion coefficient using the approximation $\mathcal{D}_Z = \langle v_Z ^2\rangle t_{\rm corr}$, where $t_{\rm corr}$ is the correlation time of the vertical velocity fluctuations, $v_Z$ \citep{fromang06}. 
In practice, we use the meridional velocity $v_\theta$ rather than $v_Z$. The quantity $\langle v_Z ^2\rangle$ represents the ensemble and time average of the mean velocities calculated at all grid cells at the cylindrical radii listed below. Using this definition for the diffusion coefficient implicitly assumes that the turbulence properties are uniform at all heights for each value of $R$.

\begin{figure}
\centering
\epsscale{1.1}
\plotone{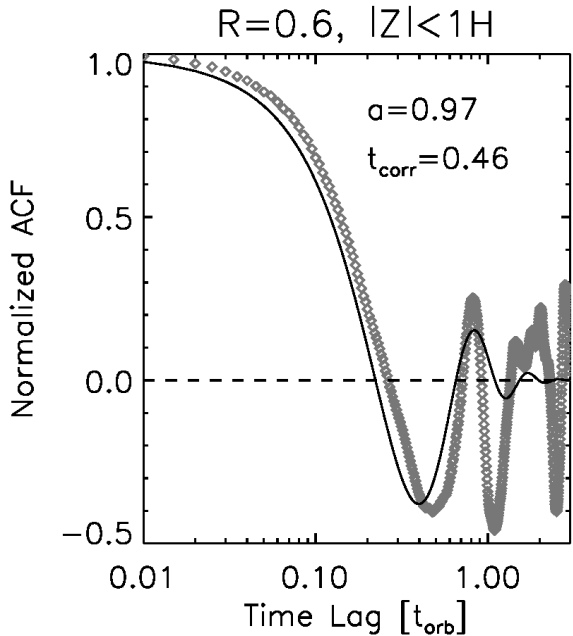}
\caption{The normalized autocorrelation function (ACF) at $R=0.6$ from the planet run with diamond symbols, and the fit obtained with Equation (\ref{eqn:tcorr}) with solid curves.}
\label{fig:acf}
\end{figure}

In obtaining an estimate for $t_{\rm corr}$, we generate a time series for $v_\theta$ at various radii in the disk ($R=0.3,0.4,0.5,0.6,$ and 0.7), within $|Z| < 1H$.
We then compute the autocorrelations of the time series and fit the computed autocorrelations with a form of 
\be
\label{eqn:tcorr}
S(t) = [(1-a) + a \cos(2 \pi \omega t)] e^{-t/t_{\rm corr}},
\en
following \citet{nelson10}. 
Here, $a$ indicates the relative strength of the sinusoidal feature in the autocorrelation function, $\omega$ is the frequency associated with the sinusoidal component, and $t_{\rm corr}$ is the correlation time.
Figure \ref{fig:acf} shows an example of the fit at $R=0.6$.
For this example, the correlation time is $0.46~t_{\rm orb}$ or $0.99~\Omega^{-1}$.
While it varies over radius, note that the correlation time is of order of $\Omega^{-1}$, which is longer than the ones obtained for MHD turbulence in protoplanetary disk simulations \citep[$\sim0.1\Omega^{-1}$;][]{fromang06,yang09,nelson10}. We suspect that this longer correlation time, combined with the prominent oscillatory component displayed by the autocorrelation function, may arise because of the contribution of the coherent radially-alternating flows that arise in the simulations as described above, such that the resulting flow is a mixture of coherent vertical motion and smaller scale turbulence.
The correlation time obtained for various radii is listed in Table \ref{tab:results}. 

With the vertical velocity fluctuation that is $5 - 11~\%$ of the local sound speed, the vertical diffusion coefficient is $(0.4 - 2.3)\times 10^{-4}$ in natural units between $R=0.3$ and $R=0.7$.
Then, the vertical mixing time $t_{\rm mix}$ can be estimated as
\be
\label{eqn:tmix}
t_{\rm mix} = H^2 /\mathcal{D}_Z.
\en
Using the $\alpha$ prescription
\be
\alpha_{\rm diff} = \mathcal{D}_Z /(c_s H),
\en
the vertical diffusion rate is characterized by values of $\alpha_{\rm diff}$ in the range $\alpha_{\rm diff} = (0.2 - 1.2) \times 10^{-2}$, which is comparable to or greater than the Reynolds stress measured at these radii (see Figure \ref{fig:evol}).

%table 1
\begin{deluxetable*}{lccccccccccc}
\tablecolumns{15}
\tabletypesize{\small}
\tablecaption{Planet Run Results \label{tab:results}}
\tablewidth{0pt}
\tablehead{
\colhead{$R$} & 
\colhead{$\langle v_\theta^2 \rangle$} &
\colhead{$\langle v_\theta^2 \rangle^{1/2} / \langle c_s \rangle$} &
\colhead{$t_{\rm corr}$} & 
\colhead{$t_{\rm corr}$} & 
\colhead{$\mathcal{D}_Z$} & 
\colhead{$\alpha_{\rm diff}$} & 
\colhead{$t_{\rm mix}$} &
\colhead{$\langle \rho_g \rangle$} &
\colhead{$(9/4)\lambda$} &
\colhead{$s_{\rm mix}$} \\
\colhead{} & 
\colhead{} & 
\colhead{} &
\colhead{($t_{\rm orb}$)} & 
\colhead{($\Omega^{-1}$)} & 
\colhead{} &
\colhead{} &
\colhead{($t_{\rm orb}$)} &
\colhead{(${\rm g~cm}^{-3}$)} & 
\colhead{(cm)}  &
\colhead{(cm)} 
}
\startdata
0.3 & $3.08\times10^{-5}$ & 0.050  & 0.43 & 2.62 & $8.32\times10^{-5}$ & 0.0065 & 0.62 & $8.05\times10^{-10}$ & 2.49 & 4.63 \\
0.4 & $3.15\times10^{-5}$ & 0.059 & 0.55 & 2.17 & $1.09\times10^{-4}$ & 0.0075 & 0.90 & $4.32\times10^{-10}$ & 4.64 & 3.20 \\
0.5 & $2.26\times10^{-5}$ & 0.055 & 0.51 & 1.44 & $7.24\times10^{-5}$ & 0.0043  & 2.06 & $3.36\times10^{-10}$ & 5.96 &  1.75 \\
0.6 & $8.08\times10^{-5}$  &  0.112 & 0.46 & 0.99 & $2.34\times10^{-4}$ &  0.0124 & 0.95 & $2.16\times10^{-10}$ & 9.28 & 3.99 \\
0.7 & $1.67\times10^{-5}$ & 0.054 &  0.41  & 0.70& $4.30\times10^{-5}$ & 0.0020 &  7.34 & $2.30\times10^{-10}$ & 8.71 & 0.82 
\enddata
\tablenotetext{a}{For the quantities in the brackets $\langle~\rangle$, we average them over azimuth within $|Z| < 1H$.}
\end{deluxetable*}

Since we are interested in the maximum size of particle that can be vertically mixed by the SWI-induced turbulence, we calculate the settling time of particles and compare it with the vertical diffusion time.
The settling time can be obtained from the terminal velocity by equating the drag force and the vertical component of the stellar gravity.
The drag force depends on the size of the particle $s$, relative to the mean free path of the gas $\lambda \equiv \mu m_{\rm H} / \rho_g \sigma$, where $\mu=2.4$ is the mean molecular weight of gas, $m_{\rm H}$ is the mass of a Hydrogen atom, $\rho_g$ is the gas mass density, and $\sigma = 2\times10^{-15}~{\rm cm}^2$ \citep{chapman70} is the molecular collisional cross section.
When $s \leq (9/4)\lambda$, the Epstein law is applicable and the drag force $F_D$ can be written as 
\be
\label{eqn:drag_ep}
F_D = {4 \over 3} \pi s^2 \rho_g v_g v_{\rm th},
\en
where $v_g$ is the gas velocity and $v_{\rm th} = \sqrt{8/\pi}c_s$ is the thermal velocity of gas molecules. 
When $s \geq (9/4)\lambda$, we can apply the Stokes law and the drag force $F_D$ can be written as 
\be
\label{eqn:drag_st}
F_D = {1 \over 2} \pi s^2 \rho_g v_g^2 C_D,
\en 
where $C_D$ is the drag coefficient.
We use $C_D = 24 Re^{-1}$ \citep{probstein69,whipple72}, since the Reynolds number $Re$ is measured to be always smaller than unity in the region and for the particle sizes of interest.
Here, $Re \equiv 2 s \rho_g v_g / \eta$ and $\eta \equiv (1/2) \rho_g v_{\rm th} \lambda$ is the gas molecular viscosity.

By setting $F_D = m_s \Omega_K^2 z$, where $m_s = (4/3) \pi s^3 \rho_s$ is the mass of a solid particle and $\rho_s = 3~{\rm g~cm^{-3}}$ is the internal density of the dust particles, one can obtain the settling time $t_{\rm settle}= z / v_g$.
For the Epstein regime, this gives
\be
\label{eqn:tsettle_ep}
t_{\rm settle} = {1 \over s} {\rho_g \over \rho_s} {v_{\rm th} \over \Omega_K^2},
\en
whereas for the Stokes regime,
\be
\label{eqn:tsettle_st}
t_{\rm settle} = {9 \over 4} {\lambda \over s^2} {\rho_g \over \rho_s} {v_{\rm th} \over \Omega_K^2}.
\en
It is immediately obvious from Equations (\ref{eqn:tsettle_ep}) and (\ref{eqn:tsettle_st}) that the equations result in the same settling time when $s = (9/4)\lambda$.
Also, note that Equations (\ref{eqn:tsettle_ep}) can be written as $t_{\rm settle} = 1/( t_{\rm s} \Omega_K^2)$, or $t_{\rm settle}/t_{\rm dyn} = 1/( 2 \pi t_{\rm s} \Omega_K)$, where the stopping time $t_{\rm s}$ for the Epstein regime is defined by $t_{\rm s} = (\rho_s  s )/(\rho_g v_{\rm th})$ \citep{weidenschilling77} and $t_{\rm dyn} = 2\pi/\Omega_K$.

Then, the maximum particle size $s_{\rm mix}$ that is significantly mixed by the turbulence (i.e., $t_{\rm mix} \sim t_{\rm settle}$) can be computed by setting Equation (\ref{eqn:tmix}) to either Equation (\ref{eqn:tsettle_ep}) or (\ref{eqn:tsettle_st}).
This leads to
\be
s_{\rm mix}=\begin{cases}
\sqrt{\dfrac{8}{\pi}} \dfrac{\rho_g}{\rho_s} \dfrac{\mathcal{D}_Z}{c_s H} H &
\text{for the Epstein regime}, \\
\left(\dfrac{9}{4} \sqrt{\dfrac{8}{\pi}} \dfrac{\rho_g}{\rho_s} \dfrac{\mathcal{D}_Z}{c_s H} H \lambda \right)^{1/2}&
\text{for the Stokes regime}. \\
\end{cases}\label{eqn:smix}
\en
As seen in Table \ref{tab:results}, in the interval $0.3 \leq R \leq 0.7$, we find that solid particles $s \sim 0.8 - 4.6$~cm in size can be vertically mixed within the first pressure scale height.

%figure 11

We note that the diffusion coefficient $\mathcal{D}_Z$ obtained from our simulation is for gas molecules.
For solid particles that are not perfectly coupled to gas, the diffusion coefficient is $\mathcal{D}_{Z,s} \simeq \mathcal{D}_Z/(1 + St^2)$ \citep{youdin07}, where $St$ is the Stokes number.
In the disk assumed here, however, centimeter-sized particles have $St \ll 1$ within the first scale height throughout the entire inner disk so $\mathcal{D}_{Z,s} \simeq \mathcal{D}_Z$.
The correlation time, vertical diffusion coefficient, the mixing time, and the maximum particle size that can be vertically mixed by the turbulence obtained for the planet run are summarized in Table \ref{tab:results}.

The above calculations relating to particle mixing assume that the SWI generates homogeneous hydrodynamic turbulence at each radius considered, but as discussed already the simulations also show strong evidence for there being coherent large scale vertical motions that might also contribute to lofting particles away from the midplane. 
Particles can only be advected within the coherent vertical flows if the vertical drag force that they feel due to the moving gas exceeds the vertical force of gravity. 
The maximum particle size for which this is true can be estimated by equating these forces -- $F_D = m_s \Omega_K^2 z$ --, assuming that the vertical gas velocity is equal to its characteristic value measured in the simulations, which is $\sim 10 \%$ of the local sound speed. Evaluating this at $R=0.6$, corresponding to 3~au in physical units, we obtain a maximum particle size of $s_{\rm max}\sim 1$~cm that can be lofted, consistent with the value obtained from the calculations above within a factor of few.

\subsection{Implications for the Growth of Large Asteroids and Terrestrial Planet Embryos}

Through a combination of the streaming instability producing planetesimals with sizes up to $\sim$ few $\times 100$~km \citep{youdin05,johansen07,johansen15} and subsequent pebble accretion onto the larger planetesimals \citep{johansen10,ormel10,lambrechts12,morbidelli12}, it seems plausible that, at 5~au in a proto-solar nebula-like gaseous disk, planetesimals can grow to become 10~Earth-mass ($M_\earth$) solid cores within about one million years \citep{bitsch15,levison15,morbidelli15}.
This presumably allows sufficient time for such cores to accrete a substantial amount of gas to eventually form gas giants, assuming that another million or so years is required for completion of the gas accretion phase \citep[e.g.,][]{pollack96,movshovitz10}, and that protoplanetary disks around T Tauri stars dissipate in a few million years \citep{haisch01,hernandez07,mamajek09}. 
While the details of how gas giant planets accrete their gaseous envelopes remain uncertain, it is clear that they must form before protoplanetary disks lose a significant fraction of their gas content (but see \citealt{tanigawa16}).

In the Solar System, Jupiter is generally thought to have been the first planet that finished accretion.
By the time Jupiter has formed, it is commonly assumed that terrestrial bodies have grown to become Moon or Mars mass embryos (see review by \citealt{dauphas11} and \citealt{morbidelli12review} and references therein).
Planetary embryos then complete their growth during the gas-free debris phase through collisions with planetesimals and/or other embryos, which takes several tens to hundreds million years after gas has removed.

More recently, models of the growth of terrestrial planet embryos and large asteroids through the accretion of pebbles and/or chondrules that are continuously generated over multi-Myr time scales have been presented \citep{johansen15,levison15}. 
In the work of \cite{johansen15} it was shown that the formation of planetesimals via high resolution streaming instability simulations leads to the formation of bodies with sizes up to $\sim 100$~km, and that subsequent chondrule accretion onto these bodies over time scales of $\sim 3$~Myr is needed to explain the presence of large asteroids such as Ceres and Vesta in the asteroid belt. 
Similarly, a model of planetary embryo growth that involves only the collisional accretion of planetesimals formed by the streaming instability was shown to be unable to form Mars-size embryos within the required time frame, and that the addition to the model of chondrule accretion over Myr time scales led to a dramatic increase in the efficacy of embryo formation. 
\cite{levison15} have presented a model in which embryos in the terrestrial planet region grow through pebble accretion over Myr time scales, with the pebbles being continuously generated during this time period, and show that such a model is able to explain certain features of the terrestrial planet system. 
The model introduces the outer solar system planets at the end of the embryo formation epoch, and subsequent evolution through giant impacts leads to final planetary systems that appear to be a good fit to the basic structure of the inner solar system, including a small mass for Mars.  

The results of our present work suggest that a Jupiter-mass planet forming within the first few Myr in a protoplanetary disk can produce turbulence and vertical stirring of solid particles interior to its orbit. 
The influence seems to be quite significant, such that solid particles with sizes up to several centimeters can be vertically well mixed. 
While pebble/chondrule accretion takes advantage of the large accretion cross section produced by aerodynamic drag, which can be orders of magnitude larger than the geometric cross section of the target planetesimals, for solid particles having a scale height ($H_p$) that is larger than the size of the planetesimal Hill radius ($r_H$) the accretion rate will be reduced by a factor of $r_H/H_p$ \citep{lambrechts12}. 
For example, assuming the planetesimal mass of $10^{24}$~g with which $r_H/R \sim 5.5\times10^{-4}$, the accretion rate for particles that have same scale height as the gas ($H/R  = H_p/R = 0.05$) will therefore drop by a factor of $\sim 90$.
This probably results in too long a timescale to form planetary embryos and large asteroids through pebble accretion within the typical lifetime of protoplanetary disks.
While this argument considers the decrease in the number density of pebbles near the midplane only, the larger relative velocity between pebbles and planetesimals as well as the non-zero orbital inclination and eccentricity of planetesimals that might arise in the SWI-driven turbulence can make the situation even worse.
Thus, {\it if} pebble accretion is the dominant process by which terrestrial planet embryos and large asteroids gain their mass, the requirement for efficient pebble accretion may put a time and/or space constraint on giant planet formation: planetary embryos have to form before a gas giant forms in the outer disk, otherwise the strong stirring induced by the SWI will significantly drop the chondrule/pebble accretion efficiency. 
It is interesting to note that isotope measurements of Martian meteorites indicate that Mars, as a  stranded planetary embryo \citep[e.g.,][]{chambers98}, grew rapidly within about 2~Myr after the birth of the solar system \citep{dauphas11b}, and then halted accretion. It is interesting to speculate that the formation of Jupiter at or close to its current location may have occurred at around this time, making pebble accretion onto Mars inefficient and effectively halting its growth and that of the large asteroids.

With our initial density profile the disk is optically thick in the midplane.
The main body of the disk therefore behaves fully adiabatically, confining the SWI towards the midplane.
As the disk evolves the optical depth of the disk will decrease by accretion of material and/or particle growth and settling, and therefore the disk will behave more isothermally. 
This will allow the SWI to operate in broader disk regions in height and possibly strengthen the spiral arms with less resistance from gas pressure, although the turbulence level and the significance of vertical mixing have to be further examined with relevant disk models.
If the level of turbulence via the SWI remains relatively constant, regardless of the thermal properties of the disk, or decreases over time, this implies that $s_{\rm mix}$ will decrease as the disk loses its mass over time (see Equation \ref{eqn:smix}).
In this case, it is possible that planetesimals and/or planetary embryos resume accreting pebbles during the later evolution as far as the disk still has some pebbles that survived rapid migration.
On the other hand, when the disk becomes optically thin the vertical shear instability and/or the MRI could operate down to the midplane and provide some turbulence that may limit pebble accretion efficiency.

If a giant planet migrates over a significant distance to reach its final position, as in the ``Grand Tack'' model proposed for the early Solar System \citep[e.g.,][]{walsh11}, the SWI could have a significant effect over a broader range of disk radii than otherwise, because the growth timescale of the instability is orders of magnitude shorter than the migration time scale.

It might be possible that gas giants form as early as during the time when their host disks still gain mass from their natal clouds, as recent observation of HL Tau may suggest \citep{alma15}. Whether or not such an early giant planet formation is common is still an open question.
If giant planets form at such an early time, the disk may be left with a very narrow window in time to grow terrestrial bodies in the disk if pebble accretion is the dominant mechanism.

\subsection{Dependence of SWI on Planetary Mass}

We have focused on the SWI arising from the spiral waves excited by a giant planet in the inner regions of a protoplanetary disk in this paper.
One may wonder if there is a mass requirement for triggering the SWI, and how the strength of the SWI varies with planet mass. 

First, given that the instability involves energy from the spiral waves being transferred into inertial modes, it is clear that stronger spiral waves will generate higher amplitudes for unstable inertial modes.
In the nonlinear saturated state, we therefore expect that the resulting hydrodynamic turbulence will be more vigorous for larger mass planets. 

Second, our general picture of how the instability operates is that a protoplanetary disk hosts the full spectrum of inertial modes that are present at low amplitude.
When a source of spiral waves (e.g., a planet) is present, it excites waves with a range of azimuthal mode numbers and Doppler-shifted frequencies.
Then, the inertial modes that are resonant with spiral waves with specific azimuthal mode numbers are able to grow due to the periodic forcing. 
In the case of a giant, gap forming planet, we have shown that the dominant spiral modes that propagate in the inner disk are the $m=2$ and $m=3$ modes, presumably because these are excited at ILRs that lie outside of the low density gap region.  
As the planet mass is lowered, however, the gap depth reduces and we expect that higher values of $m$ will become prominent, with the amplitudes of waves associated with lower $m$ values decreasing in proportion to the planet mass. 
The increasing prominence of the higher $m$ spiral modes may then allow the SWI to operate in regions closer to the planet because, at the midplane, the SWI operates in the radial intervals $((m-2)/m)^{2/3} R_p \leq R \leq ((m-1)/m)^{2/3} R_p$ in the inner disk and in $((m+1)/m)^{2/3} R_p \leq R \leq ((m+2)/m)^{2/3} R_p$ in the outer disk. 
Again, these conditions come from the fact that, as the value of $m$ increases, the synodic period between orbiting fluid elements and the planet needs to increase to match the resonance condition with inertial modes in the disk.

Based on the two points discussed above, we expect that (1) the strength of SWI-driven turbulence will be weaker for lower mass planets; and (2) the region where the instability operates will be narrower and closer to the planet, as the planet mass decreases to the point that the $m=2$ mode is no longer strong enough to induce the SWI in the inner disk.

In order to support these two points, we have run additional simulations with planetary masses of $q \equiv M_p/ M_* = 10^{-3}, 5\times10^{-4}, 3\times10^{-4}, 10^{-4}, 10^{-5}$.
For these runs, we were forced to use a numerical resolution of $(N_r, N_\theta, N_\phi) = (364,72,378)$ because of the computational cost. 
This is a factor of two lower than our reference model and so thus the growth of the SWI is not properly captured, although we can measure spiral wave amplitudes. 

\begin{figure}
\centering
\epsscale{1.15}
\plotone{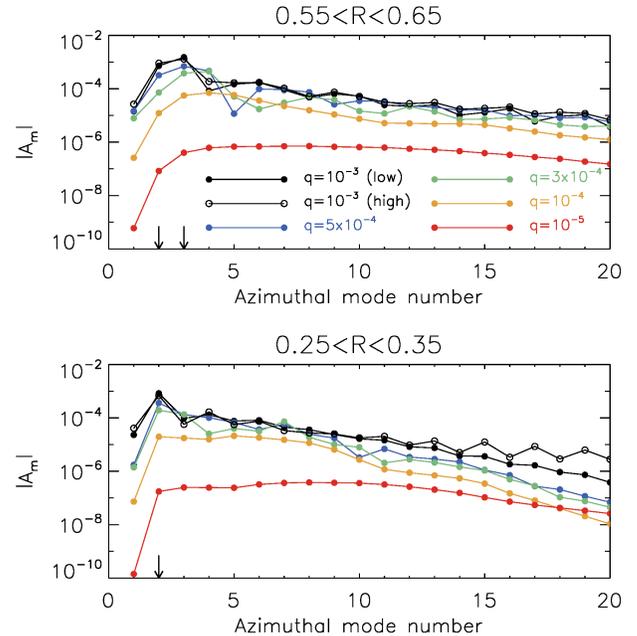}
\caption{The Fourier amplitude $|A_m|$ for various azimuthal modes at (upper) $0.55<R<0.65$ and (lower) $0.25<R<0.35$, taken at $t=100~t_{\rm orb}$ for different planetary masses $q = M_p / M_*$ of (black) $10^{-3}$, (blue) $5\times10^{-4}$, (green) $3\times10^{-4}$, (yellow) $10^{-4}$, and (red) $10^{-5}$. Black open circles are for our reference model with the high numerical resolution of $(N_r,N_\theta,N_\phi) = (726,144,754)$, whereas all the other models run with low resolution of $(N_r,N_\theta,N_\phi) = (364,72,378)$. The arrows indicate the azimuthal mode numbers with which perturbing waves are capable of exciting the SWI at each radius bin. }
\label{fig:fpower2}
\end{figure}

The Fourier amplitudes for different planetary masses are plotted in Figure \ref{fig:fpower2} as a function of azimuthal mode number.
At both $0.55 < R < 0.65$ and $0.25 < R < 0.35$, the Fourier amplitudes are larger for more massive planets in general.
It is therefore reasonable to expect that more massive planets will produce larger energy injection rates into the inertial modes, and accordingly stronger turbulence through the SWI.
We see that the maximum amplitude shifts towards lower degree azimuthal modes as $q$ increases, presumably due to non-linear mode coupling \citep[e.g.,][]{lee16} and gap formation which suppresses the amplitudes of the high-$m$ modes that are excited close to the planet. 
This trend is particularly apparent for $q \ge 3 \times 10^{-4}$.
As the SWI is expected to operate for $m=2$ and $m=3$ perturbations at radii $0.55 < R < 0.65$ (it can also be excited by the $m=4$ mode between $0.63 \le R/R_p \le 0.65$, but this is only just contained in the radial range under consideration), and for $m=2$ modes only at radii $0.25 < R < 0.35$, we expect that the SWI operates more strongly with larger planetary masses that have most power in $m=2$ or $m=3$ in these regions ($q \ge 3 \times 10^{-4}$).
When $q=10^{-5}$, the Fourier amplitudes are relatively flat for $m \simeq 3 - 10$. 
If the SWI can operate for such a low mass planet, then we might expect that the strength of the outcome will be much weaker than for larger mass planets, and that the regions unstable to the instability will be limited to a narrow radial region around the planetary orbit because of the weakness of the $m=2$ spiral wave.
Global disk calculations performed at high resolution will be required to confirm or refute these conjectures, and to fully characterize how the SWI operates as a function of planet mass.

%figure 12

\subsection{Caveats and Future Work}

As seen in Figure \ref{fig:acf}, the ACF of the vertical velocity does not converge but oscillates around zero with finite amplitude.
This is because the vertical flows developing from the SWI not only contain repeated growth and decay of vertical motions, but also display sustained and coherent oscillations, likely due to the excitation of longer wavelength inertial waves.
Due to this apparently complex superposition of turbulence and coherent oscillations, the analysis presented in Section \ref{sec:diffusion} has to be viewed with some caution by keeping in mind the uncertainty in estimating the correlation time. 
Nonetheless, a factor of a few uncertainty in the correlation time would not change our conclusions about particle stirring in a qualitative sense.

At this point, it is probably worthwhile pointing out that some previous simulation studies, in which grains and pebbles are implemented as Lagrangian particles, indicate evidence of vertical mixing of solid particles through the SWI.
As a potential example, we note that in Figure 13 of \citet{zhu14}, where the authors placed a 0.65 Jupiter-mass planet around a solar-mass star, particles a and b -- that are about a millimeter and a centimeter in size, respectively, when the MMSN disk model and the semi-major axis of 5~au are assumed -- are vertically dispersed in the outer disk ($R > 1$ in the figure). 
In particular, the particle distributions show some wiggling morphology near the midplane, which is tempting to explain with the radially alternating vertical flows arising from the SWI.
The vertical dispersal of particles is not seen in the inner disk and we speculate that this is because of the insufficient numerical resolution to properly capture the unstable inertial modes there.
They used uniform radial grid cells, with which one scale height is resolved with only about 5 grid cells in both radial and vertical directions at $R=0.7$, as opposed to with about 15 grid cells at $R=1.5$.

In the future, more quantitative conclusions will be able to be made from high resolution gas and particle simulations, where one can obtain the vertical distribution of solid particles stirred by the SWI, both interior and exterior to the planet. Furthermore, some of the questions that we have discussed in this paper such as the outcome of the SWI as a function of planet mass, and the influence that this has on pebble accretion, can also be addressed in the context of improved disk models that contain a more sophisticated model of disk thermodynamics. And finally, hydrodynamic simulations combined with radiative transfer calculations can be used to determine the observational appearance of disks with planets, and to determine what influence the SWI has on images formed through scattered light and thermal emission.

\section{CONCLUSION}
\label{sec:conclusion}

We have presented inviscid three-dimensional global hydrodynamic simulations of protoplanetary disks in which we excite spiral waves either monochromatically using an imposed external potential or using a full planetary potential.
Using monochromatic waves, we first show that the region of the SWI operating and the properties of the unstable inertial modes show very good agreement with the predictions made by matching resonance criteria using simple dispersion relations for spiral and inertial waves.
When a Jupiter-mass planet is added in the disk, the spiral waves it launches have various azimuthal components superimposed on the dominant $m=2$ or $m=3$ modes, depending on the radial position in the disk, and
the SWI is found to grow on the order of the planet orbital time.  
When it is saturated, the SWI generates turbulence and radially alternating vertical flows that have characteristic radial and vertical length scales of about one local pressure scale height and vertical velocities of order of few tens percent of the local sound speed.
Using the alpha prescription, the associated vertical diffusion rate of gas is estimated to be characterized by $\alpha_{\rm diff} \sim (0.2-1.2)\times 10^{-2}$ in the range of radii $0.3 R_p \leq R \leq 0.7 R_p$, where $R_p$ is the semi-major axis of the planet.
At this rate, solid particles up to a few centimeters in size can be vertically mixed within the first pressure scale height.
Since protoplanetary disks are believed to remain laminar, and thus induce no or very little particle stirring, as suggested by recent magnetized wind models \citep{bai13,gressel15}, the SWI can be the mechanism controlling the degree of vertical settling of solid particles in the optically-thick regions of planet-hosting disks where other hydrodynamic instabilities are not thought to operate.

While more quantitative results will be obtained with future high resolution simulations that include particles, we conjecture that significant stirring of solid particles through the instability of the spiral waves excited by giant planets can have an influence on the formation and growth of large asteroids and terrestrial planet embryos in the inner disk, if these bodies are actively growing in the presence of a giant planet (such as Jupiter in the protosolar nebula).
In particular, if growth of these bodies proceeds mainly through the accretion of chondrules/pebbles, then they have to form before a gas giant forms in the outer disk as the stirring of solid particles is likely to significantly decrease the chondrule/pebble accretion efficiency.
Since the properties of the turbulence driven by the SWI are dependent upon the background thermal structure of the disk as well as the planetary mass, future studies that survey parameter space at high numerical resolution are needed to further examine the significance of the instability for protoplanetary disk evolution and for planet formation.

\acknowledgments

The authors thank the anonymous referee for a prompt report and helpful comments that improved the initial manuscript.
J.B. thanks Edwin Bergin, Jeffrey Fung, Wing-Kit Lee, and Zhaohuan Zhu for helpful conversations.
This research was supported in part through computational resources and services provided by Advanced Research Computing at the University of Michigan, Ann Arbor.
This work used the Extreme Science and Engineering Discovery Environment (XSEDE), which is supported by National Science Foundation grant number ACI-1053575.
The authors acknowledge the San Diego Supercomputer Center at University of California, San Diego for providing HPC resources that have contributed to the research results reported within this paper.


\begin{thebibliography}{99}

\bibitem[ALMA Partnership et al.(2015)]{alma15} ALMA Partnership, Brogan, C.L., P\'erez, L. M., et al. \ 2015, \apjl, 808, L3 

\bibitem[Artymowicz \& Lubow(1992)]{artymowicz92} Artymowicz, P., \& Lubow, S.~H.\ 1992, \apj, 389, 129 

\bibitem[Bae et al.(2016)]{bae16} Bae, J., Nelson, R. P., Hartmann, L., \& Richard, S. 2016, \apj, 829, 13

\bibitem[Bai \& Stone(2013)]{bai13} Bai, X.-N \& Stone, J. M. 2013, \apj, 769, 76

\bibitem[Benisty et al.(2015)]{benisty15} Benisty, M., Juhasz, A., Boccaletti, A., et al. 2015, \aap, 578, L6

\bibitem[Ben\'itez-Llambay \& Masset(2016)]{benitez16} Ben\'itez-Llambay, P., \& Masset, F. 2016, \apjs, 223, 11

\bibitem[Bitsch et al.(2015)]{bitsch15} Bitsch, B., Lambrechts, M., \& Johansen, A. 2015, \aap, 582, A112

\bibitem[Chambers \& Wetherill(1998)]{chambers98} Chambers, J.~E., \& Wetherill, G. W. 1998, Icarus, 136, 304

\bibitem[Chapman \& Cowling(1970)]{chapman70} Chapman, S., \& Cowling, T. G. \ 1970, The Mathematical Theory of Non-uniform Gases. An account of the Kinetic Theory of Viscosity, Thermal Conduction and Diffusion in Gases (3rd ed.; Cambridge: Cambridge Univ. Press)

\bibitem[Currie et al.(2014)]{currie14} Currie, T., Muto, T., Kudo, T., et al. 2014, \apjl, 796, L30

\bibitem[Dauphas \& Chaussidon(2011)]{dauphas11} Dauphas, N., \& Chaussidon, M.\ 2011, AREPS, 39, 351 

\bibitem[Dauphas \& Pourmand(2011)]{dauphas11b} Dauphas, N., \& Pourmand, A.\ 2011, Natur, 473, 489 

\bibitem[D'Angelo et al.(2003)]{dangelo03} D'Angelo, G., Henning, T., \& Kley, W. \ 2003, \apj, 599, 548

\bibitem[de Val-Borro et al.(2006)]{devalborro06} de Val-Borro, Edgar, R. G., M., Artymowicz, P., et al. \ 2006, \mnras, 370, 529

\bibitem[Fromang \& Papaloizou(2006)]{fromang06} Fromang, S., \& Papaloizou, J.\ 2006, \aap, 452, 751 

\bibitem[Fromang \& Papaloizou(2007)]{fromang07} Fromang, S., \& Papaloizou, J.\ 2007, \aap, 468, 1

\bibitem[Fung \& Dong(2015)]{fung15} Fung, J., \& Dong, R.\ 2015, \apjl, 815, L21

\bibitem[Fung \& Chiang(2016)]{fung16} Fung, J., \& Chiang, E.\ 2016, , 815, L21

\bibitem[Garufi et al.(2013)]{garufi13} Garufi, A., Quanz, S. P., Avenhaus, H., et al. 2013, \aap, 560, A105

\bibitem[Garufi et al.(2016)]{garufi16} Garufi, A., Quanz, S.P., \& Schmid, H. M. 2016, \aap, 588, A8

\bibitem[Goldreich \& Tremaine(1979)]{goldreich79} Goldreich, P., \& Tremaine, S. \ 1979, \apj, 233, 857

\bibitem[Goldreich \& Tremaine(1980)]{goldreich80} Goldreich, P., \& Tremaine, S. \ 1980, \apj, 241, 425

\bibitem[Goodman(1993)]{goodman93} Goodman, J.\ 1993, \apj, 406, 596 

\bibitem[Grady et al.(2013)]{grady13} Grady, C. A., Muto, T., Hashimoto, J., et al. 2013, \apj, 762, 48

\bibitem[Gressel et al.(2015)]{gressel15} Gressel, O., Turner, N.~J., Nelson, R.~P., \& McNally, C. P.\ 2015, \apj, 801, 84

\bibitem[Haisch et al.(2001)]{haisch01} Haisch, K. E., Lada, E. A., \& Lada, C. J. 2001, \apj, 553, L153

\bibitem[Hayashi(1981)]{hayashi81} Hayashi, C.\ 1981, Progress of Theoretical Physics Supplement, 70, 35

\bibitem[Hern\'andez et al.(2007)]{hernandez07} Hern\'andez, J., Hartmann, L., Megeath, T., et al. 2007, \apj, 662, 1067

\bibitem[Horn et al.(2012)]{horn12} Horn, B., Lyra, W., Mac Low, M.-M., \& S\'andor, Z. \ 2012, \apj, 750, 34

\bibitem[Hubeny(1990)]{hubeny90} Hubeny, I.\ 1990, \apj, 351, 632

\bibitem[Johansen et al.(2007)]{johansen07} Johansen, A., Oishi, J. S., Mac Low, M.-M., et al. 2007, Natur, 448, 1022

\bibitem[Johansen \& Lacerda(2010)]{johansen10} Johansen, A., \& Lacerda, P.\ 2010, \mnras, 404, 475

\bibitem[Johansen et al.(2015)]{johansen15} Johansen, A., Mac Low, M.-M., Lacerda, P., \& Bizzarro, M. 2015, SciA, 1, 1500109

\bibitem[Juh\'asz et al.(2015)]{juhasz15} Juh\'asz, A., Benisty, M., Pohl, A., et al. \ 2015, \mnras, 451, 1147 

\bibitem[Lambrechts \& Johansen(2012)]{lambrechts12} Lambrechts, M., \& Johansen, A. 2012, \aap, 544, A32

\bibitem[Lee(2016)]{lee16} Lee, W.-K. \ 2016, arXiv:1604.08941 

\bibitem[Levison et al.(2015)]{levison15} Levison, H.~F., Kretke, K.~A., \& Duncan, M. J.\ 2015a, Natur, 524, 322 

\bibitem[Lyra et al.(2010)]{lyra10} Lyra, W., Paardekooper, S.-J., \& Mac Low, M.-M. \ 2010, \apjl, 715, L68 

\bibitem[Lyra et al.(2016)]{lyra16} Lyra, W., Richert, A.~J.~W., Boley, A., et al.\ 2016, \apj, 817, 102 

\bibitem[Mamajek(2009)]{mamajek09} Mamajek, E. E. 2009, in AIP Conf. Proc. 1158, Exoplanets and Disks: Their Formation and Diversity, ed. T. Usuda, M. Tamura, \& M. Ishii (Melville, NY: AIP), 3

\bibitem[Masset(2000)]{masset00} Masset, F.\ 2000, \aaps, 141, 165 

\bibitem[Morbidelli \& Nesvorny(2012)]{morbidelli12} Morbidelli, A., \& Nesvorny, D.\ 2012, \aap, 546, A18

\bibitem[Morbidelli et al.(2012)]{morbidelli12review} Morbidelli, A., Lunine, J. I., O'Brian, D. P., Raymond, S. N., \& Walsh, K. J. \ 2012, AREPS, 40, 251

\bibitem[Morbidelli et al.(2015)]{morbidelli15} Morbidelli, A., Lambrechts, M., Jacobson, S., \& Bitsch, B. \ 2015, Icarus, 258, 418

\bibitem[Movshovitz et al.(2010)]{movshovitz10} Movshovitz, N., Bodenheimer, P., Podolak, M., \& Lissauer, J.~J.\ 2010, \icarus, 209, 616 

\bibitem[Muto et al.(2012)]{muto12} Muto, T., Grady, C. A., Hashimoto, J., et al.\ 2012, \apjl, 748, L22 

\bibitem[Nelson \& Gressel(2010)]{nelson10} Nelson, R.~P., \& Gressel, O.\ 2010, \mnras, 409, 639

\bibitem[Nelson et al.(2013)]{nelson13} Nelson, R. P., Gressel, O., \& Umurhan, O. M. \ 2013, \mnras, 435, 2610

\bibitem[Ormel \& Klahr(2010)]{ormel10} Ormel, C. W., \& Klahr, H. H.\ 2010, \aap, 520, A43 

\bibitem[Pollack et al.(1996)]{pollack96} Pollack, J. B., Hubickyj, O., Bodenheimer, P., et al. 1996, Icarus, 124, 62

\bibitem[Probstein \& Fassio (1969)]{probstein69} Probstein, R. F., \& Fassio, F. \ 1969, AIAA, 8, 4 

\bibitem[Shakura \& Sunyaev(1973)]{shakura73} Shakura, N.~I., \& Sunyaev, R.~A.\ 1973, \aap, 24, 337 

\bibitem[Stone \& Norman(1992)]{stone92} Stone, J.~M., \& Norman, M.~L.\ 1992, \apjs, 80, 753 

\bibitem[Stolker et al.(2016)]{stolker16} Stolker, T., Dominik, C., Avenhaus, H., et al. \ 2016, arXiv:1603.00481

\bibitem[Tanigawa \& Tanaka(2016)]{tanigawa16} Tanigawa, T., \& Tanaka, H.\ 2016, \apj, 823, 48

\bibitem[Testi et al.(2014)]{testi14} Testi, L., Birnstiel, T., Ricci, L., et al. \ 2014, in Protostars and Planets VI, ed. H. Beuther et al. (Tucson, AZ: Univ. of Arizona Press), 339 

\bibitem[Walsh et al.(2011)]{walsh11} Walsh, K.~J., Morbidelli, A., Raymond, S.~N., O'Brien, D.~P., \& Mandell, A.~M.\ 2011, \nat, 475, 206 

\bibitem[Whipple(1972)]{whipple72} Whipple, F. L.\ 1972, From Plasma to Planet, ed. A. Evlius (NewYork:Wiley), 211 

\bibitem[Weidenschilling(1977)]{weidenschilling77} Weidenschilling, S.~J.\ 1977, \mnras, 180, 57 

\bibitem[Yang et al.(2009)]{yang09} Yang, C.-C., Mac Low, M.-M, \& Menou, K.\ 2009, \apj, 707, 1233 

\bibitem[Youdin \& Goodman(2005)]{youdin05} Youdin, A. N., \& Goodman, J.\ 2005, \apj, 620, 459

\bibitem[Youdin \& Lithwick(2007)]{youdin07} Youdin, A. N., \& Lithwick, Y.  2007, Icarus, 192, 588

\bibitem[Zhu et al.(2009)]{zhu09} Zhu, Z., Hartmann, L., \& Gammie, C.\ 2009, \apj, 694, 1045 

\bibitem[Zhu et al.(2014)]{zhu14} Zhu, Z., Stone, J. M., Rafikov, R. R., \& Bai, X.-N.  2014, \apj, 785, 122

\bibitem[Zhu et al.(2015)]{zhu15} Zhu, Z., Dong, R., Stone, J. M., \& Rafikov, R. R.  2015, \apj, 813, 88

\end{thebibliography}
\end{document}